\documentclass[11pt]{article}

\input epsf
\newcommand{\be}{\begin{equation}}
\newcommand{\ee}{\end{equation}}
\newcommand{\bea}{\begin{eqnarray}}
\newcommand{\eea}{\end{eqnarray}}
\newcommand{\bref}[1]{(\ref{#1})}

\newcommand{\ct}[1]{\cite{#1}}

\newcommand{\gh}[1]{{\cal #1}}
\newcommand{\expect}[1]{\left\langle \, #1 \, \right\rangle}

\makeatletter
\def\restric#1#2{{\left. #1 \right|_{#2}}}
\def\dif{{\rm d}}
\def\secteqno{\@addtoreset{equation}{section}%
\def\theequation{\thesection.\arabic{equation}}}
\def\endsecteqno{\def\theequation{\@ifundefined{chapter}%
{\arabic{equation}}{\thechapter.\arabic{equation}}}}
\newcounter{subequation}
\def\thesubequation{\alph{subequation}}
\def\sneqnarray{\stepcounter{equation}\let\@currentlabel=\theequation
\setcounter{subequation}{1}
\def\@eqnnum{{\rm (\theequation.\thesubequation)}}
\global\@eqcnt\z@\tabskip\@centering\let\\=\@eqncr\let\@@
eqncr=\@@sneqncr
$$\halign to \displaywidth\bgroup\@eqnsel\hskip\@centering
 $\displaystyle\tabskip\z@{##}$&\global\@eqcnt\@ne
 \hskip 2\arraycolsep \hfil${##}$\hfil
 &\global\@eqcnt\tw@ \hskip 2\arraycolsep
$\displaystyle\tabskip\z@{##}$\hfil
  \tabskip\@centering&\llap{##}\tabskip\z@\cr}
\def\endsneqnarray{\@@sneqncr\egroup$$\global\@ignoretrue}
\def\@@sneqncr{\let\@tempa\relax
   \ifcase\@eqcnt \def\@tempa{& & &}\or \def\@tempa{& &}
   \else \def\@tempa{&}\fi
     \@tempa \if@eqnsw\@eqnnum\stepcounter{subequation}\fi
     \global\@eqnswtrue\global\@eqcnt\z@\cr}
\makeatother

\def\AP#1#2#3{ {{\sl Ann.\,Phys.\,}(N.Y.)\,}
    {\bf  {#1}} ({#2}) {#3}}
\def\CMP#1#2#3{ {\sl Commun.\,Math.\,Phys.\,}
    {\bf  {#1}} ({#2}) {#3}}
\def\CCG#1#2#3{ {\sl Class.\,Quantum\,Grav.\,}
    {\bf  {#1}} ({#2}) {#3}}
\def\IJMPA#1#2#3{ {\sl Int.\,J.\,Mod.\,Phys.\,}
    {\bf A{#1}} ({#2}) {#3}}
\def\JMP#1#2#3{ {\sl J.\,Math.\,Phys.\,}
    {\bf  {#1}} ({#2}) {#3}}

\def\NC#1#2#3{ {\sl Nuovo\,Cim.\,}
    {\bf  {#1}} ({#2}) {#3}}
\def\NPB#1#2#3{ {\sl Nucl.\,Phys.\,}
    {\bf B{#1}} ({#2}) {#3}}

\def\PLB#1#2#3{ {\sl Phys.\,Lett.\,}
    {\bf B{#1}} ({#2}) {#3}}
\def\PR#1#2#3{ {\sl Phys.\,Rep.\,}
    {\bf  {#1}} ({#2}) {#3}}
\def\PRL#1#2#3{ {\sl Phys.\,Rev.\,Lett.\,}
    {\bf  {#1}} ({#2}) {#3}}
\def\PRD#1#2#3{ {\sl Phys.\,Rev.\,}
    {\bf D{#1}} ({#2}) {#3}}

\newcommand{\norm}[2]{\mbox{N}_{#1}\left[#2\right]}

\newcommand{\vgl}[1]{eq.(\ref{#1})}
\newcommand{\gn}{ ghost number }
\newcommand{\dr}{\raise.3ex\hbox{$\stackrel{\leftarrow}{\delta }$}}
\newcommand{\dl}{\raise.3ex\hbox{$\stackrel{\rightarrow}{\delta}$}}
\newcommand{\cA}{{\cal A}}

\newcommand{\cG}{{\cal G}}

\newcommand{\cI}{{\cal I}}

\newcommand{\cM}{{\cal M}}

\newcommand{\cO}{{\cal O}}
\newcommand{\cP}{{\cal P}}

\newcommand{\cS}{{\cal S}}

\newcommand{\gras}[1]{\epsilon_{#1}}
\newcommand{\ihbar}{\frac{i}{\hbar}}
\newcommand{\ddr}{\raise.3ex\hbox{$\stackrel{\leftarrow}{d}$}}
\newcommand{\ddl}{\raise.3ex\hbox{$\stackrel{\rightarrow}{d}$}}

\def\gtwid{\raise.3ex\hbox{$>$\kern-.75em\lower1ex\hbox{$\sim$}}}
\def\ltwid{\raise.3ex\hbox{$<$\kern-.75em\lower1ex\hbox{$\sim$}}}
\secteqno

\begin{document}

\begin{titlepage}

\begin{flushright}
                   Preprint KUL-TF-96/4 \\
                   NIKHEF-96-009\\
                   hep-th/9603012\\
                   March 96 \\
\end{flushright}

\vfill

\begin{center}

{\LARGE\bf The BPHZ renormalised BV master equation and \\
Two-loop Anomalies in Chiral Gravities}

\vskip 20.mm

{\bf F.\,De Jonghe$^{a,1}$, J.\,Par\'{\i}s$^{b,2}$ and W.\,Troost$^{b,3}$}\\

\vskip  1cm

{\it $^a$ NIKHEF/FOM, Postbus  41882, 1009 DB Amsterdam, The Netherlands} \\

\vskip 3mm

{\it $^b$ Instituut voor Theoretische Fysica, K.U.Leuven \\
Celestijnenlaan 200D, B-3001 Leuven, Belgium}
\\[0.3cm]

\end{center}

\vfill

\begin{center}

{\bf Abstract}

\end{center}

\begin{quote}
\small
Anomalies and BRST invariance are governed, in the context of Lagrangian
Batalin-Vilkovisky quantization, by the master equation, whose classical
limit is $(S, S)=0$. Using Zimmerman's normal products and the BPHZ
renormalisation method, we obtain a corresponding local quantum operator
equation, which is valid to all orders in perturbation theory. The
formulation implies a calculational method for anomalies
that is useful also outside
this context and that remains completely within regularised perturbation
theory. It makes no difference in principle whether the anomaly appears at
one loop or at higher loops.
The method is illustrated by computing the one- and two-loop
anomalies in chiral $W_3$ gravity.

\end{quote}

\vspace{2mm}

\vfill

\hrule width 5.cm

{\small \noindent $^1$ E-mail : t54@nikhef.nl \\
\noindent $^2$ E-mail: jordi.paris@fys.kuleuven.ac.be \\
\noindent $^3$ Onderzoeksleider N.F.W.O., Belgium\\
\hspace*{5pt} E-mail: walter.troost@fys.kuleuven.ac.be}

\normalsize

\end{titlepage}

\setcounter{page}{2}

%%%%%%%%%%%%%%%%%%%%%%%%%%%%%%%%%%%%%%%%%%%%%%%%%%%%%%%%
\section{Introduction}
%%%%%%%%%%%%%%%%%%%%%%%%%%%%%%%%%%%%%%%%%%%%%%%%%%%%%%%%
\hspace{\parindent}%
The most comprehensive
method for treating gauge theories is at present the
method of Batalin and Vilkovisky (BV). It unifies in one framework
the description of gauge invariance,
the selection of physical observables,
the gauge fixing, BRS methods,
the introduction of ghosts (and antighosts, if wanted), etc.
All this is achieved irrespective of whether the gauge algebra is a
relatively simple one (as for nonabelian gauge theories) or a more
complicated one, closing only upon using the equations of motion (a BRST
charge that is nilpotent only on shell), as often occurs in
supergravity and in string theories.

Whereas over the years the quantum BRS treatment of gauge theories
\ct{Becchi} has
met with many successes \cite{Piguet and Rouet PR},
the quantum theory in the BV scheme is less developed.
Indeed, the remarks above strictly apply to the classical theory.
For many considerations, including some quantum questions,
like the classifications of one loop anomalies for example,
one practically needs use {\em only} of the classical theory:
interesting results follow by considering that in the
quantum theory the anomalies have to satisfy a certain condition
and by solving its classical version (the Wess-Zumino consistency condition).
A full quantum analysis however requires a more careful treatment.
It is rather easy to give {\em formal} arguments, and in fact they were
already given in the founding papers on the subject \cite{BV};
but when one goes
deeper into the matter, taking seriously the occurrence of
the well-known divergences, one realises that the literature is defective.
This paper addresses this problem head-on.

In the BV setup, the gauge
invariance at the classical level is expressed by the so-called master
equation, $(S,S)=0$. Here $S$ is the extended classical action,
which consists of the classical action (including ghosts and antighosts),
source terms for the BRS-transformation
laws, and some supplementary terms if the gauge
algebra is open. On the quantum level, BRST invariance can be transcribed
as $(\Gamma,\Gamma)=0$, where $\Gamma$ is the generating functional of
the one-particle irreducible graphs, or the effective action,
evaluated in the presence of the above mentioned source terms. Anomalies
consist then in a violation of this quantum equation.
In the BRS context, this question was developed making extensive use of
the formulation of the quantum theory on the level of the effective
action. This development can of course be transcribed to the BV formalism.
However, the question arises also
how this translates in the quantum theory to the {\em local}
level, i.e. to the level of the action.
Formal manipulations modify  the
classical master equation into $(S_q,S_q)=2 i \hbar\Delta S_q$,
where the quantum action possibly contains terms that are explicitly
of higher order in $\hbar$,  $S_q=S+ \gh O(\hbar)$ and $\Delta$ is
a second order functional differential operator where
the functional derivatives are at the {\em same} space-time point. If
$S_q$ is a local action, this operation leads to
$\delta(0)$, it is ill-defined. The source of the problem is of course
the same as always in local quantum field theory, namely the coincidence of
space-time points. A regularisation scheme should solve it.

Efforts in this direction include \cite{anombv} where the method of Pauli
and Villars has been applied to this question. This proposal works well on
the one loop level, but it is not very clear how to extend this type
of calculation to higher loops. In a less conventional approach,
the use of non-local regularisation was suggested \cite{j95,jw95},
providing a method valid also for
higher order loops. In \ct{Tonin}, an analysis of the quantum aspects of
BV using dimensional regularization is presented.

The interest in a full local quantum treatment is not a purely conceptual one.
It is also of importance for the question of {\it anomalies appearing
in higher loops}. Indeed, in some theories,
contrary to the situation in nonabelian gauge
theories \cite{abbj}, anomalies can arise at the multiple loop
level. This is notably the case in extended theories of two dimensional
gravity, so-called $W$-gravities. The presence of a two-loop anomaly in
$W_3$-gravity, for example, is ascertained through the use of the operator
product expansions of two dimensional conformal field theory. Whereas this
is certainly a valid method, it leaves a gap. Namely, rather than deriving
these operator product expansions from a proper treatment of the
renormalised quantum field theory that underlies them, their
validity is mostly {\em assumed}, based on algebraic and symmetry
considerations. Clearly, a complete treatment of higher loop
anomalies in renormalised perturbative quantum field theory is lacking.

The purpose of the present paper
is to formulate and derive the basic local equations governing
the BV method on the level of renormalised perturbation theory.
On a more practical level, we will provide a proper method,
always within conventional renormalised perturbation theory, to compute
higher loop anomalies. We will use the method of Bogoliubov, Parasiuk,
Hepp and Zimmerman (BPHZ) to achieve this purpose. Not only is this
method of impeccable standing, but also we will find that it is in fact
well-suited to put the quantum master equation of the BV formalism on a
solid foundation. In fact, that it is suited for computing anomalies
(in global symmetries) was already realised long ago (see \cite{bron}).
Treatments of non-abelian gauge theories, using the same operator methods,
on the level of the effective action, have of course played an important
role in firmly establishing their renormalisability
(after the dimensional regularisation treatment
of~\cite{tHooft en Veltman}). An overview of this and other developments
that goes together well with the BV scheme can be found
in~\cite{boek Piguet en Sorella}%
\footnote{This book is also useful for the  reader who needs more
background material on renormalisation and symmetries in field theory
than can be provided in this paper.}.

This paper should not be read as a complete account of the way the BV
formalism and the BPHZ method work together to provide a full account, in
renormalised quantum field theory, of all possible higher loop anomalies.
Rather it should be viewed as the onset of a program. It does provide the
basic equations and explicit examples. Working out the detailed
consequences in the most general situations is left for the future.
As far as the method to compute BRST anomalies is concerned, it can be
stripped of its BV dressing and applied outside this context as well.

We have organized the paper as follows. In section~\ref{BV} we recall
the basics of the
Batalin-Vilkovisky method, the relationship of the quantum master equation
with the occurrence of anomalies, and the Zinn-Justin equation,
thus setting
the stage for a more precise formulation of what we want to achieve.
Section~\ref{BPHZ} starts with an extensive discussion of
renormalisation schemes in the BV setting, thus argueing for the use of
the BPHZ method. The basics of this method, including the
manifestation of anomalies (in particular the chiral anomaly) are then
recalled, and some aspects of the massless case are presented from an angle
that complements the existing literature. Section~\ref{BVandBPHZ}
combines the previous developments into a general derivation of the master
equation in the BPHZ framework.
The corresponding treatment for the concrete case of $W_2$-gravity in
section~\ref{W2} serves  as an
illustration of the more abstract treatment of the previous sections,
and also as a warm-up for the treatment of $W_3$ gravity in
section~\ref{W3} wherein, after quickly giving the one-loop
anomaly, we come to the complete computation of the two-loop anomaly
of $W_3$. Three appendices collect some formulas on BPHZ renormalized
one-loop integrals for $2D$ chiral gravities,
supply the complete one-loop anomaly for $W_3$, and
illustrate the BPHZ treatment of the ghost number anomaly.

\section{Batalin-Vilkovisky Lagrangian quantisation}\label{BV}

\hspace{\parindent}%
In this section, we recall the vocabulary of the
Batalin-Vilkovisky (BV) formulation of gauge theories \cite{BV},
including the formulation of anomalies on a formal quantum level.
More detailed reviews on the BV scheme, reflecting the point of view of
the present authors, can
be found in \cite{JQS,WT,Frank}. Other recommended reading on BV is
\cite{Henneaux}.

\subsection{Basic concepts}

\hspace{\parindent}%
The BV scheme combines BRS quantisation \cite{brs} and
its extension to open algebras \cite{Open} with the observation
--originally due to Zinn-Justin (see e.g. chapter 19 of \cite{ZJbook})--
that it is convenient to introduce sources for the BRST transformation
rules during the renormalization process.
With every field%
\footnote{$\Phi^A$ is the complete set of fields in the theory: gauge fields,
matter fields, ghosts, antighosts, \ldots}
 $\Phi^A$ one associates
an antifield
$\Phi^*_A$ of opposite Grassmann parity%
\footnote{De Witt notation is assumed throughout the paper: the index $A$
is used to indicate the different fields, their components, {\em and} the
space-time point on which they depend (unless it is explicitly displayed).
In this way, a summation over $A$ includes not only discrete summations,
but also integration over space-time. \label{deWitt-conventie}
The derivatives are left and right functional derivatives}.
Fields and antifields are
canonically conjugated w.r.t.\ the {\it antibracket} defined by
\be
   (F , G) = \frac{\dr F}{\delta \Phi^A} \frac{\dl G}{\delta \Phi^*_A} -
 \frac{\dr F}{\delta \Phi^*_A} \frac{\dl G}{\delta \Phi^A}.
\label{antibracket}
\ee
The ghost number of a field and that of its antifield add up to $-1$.

The BRST invariant, classical theory is then described by an
{\it extended action} $S(\Phi,\Phi^*)$, formed by adding several pieces:
the classical action, a term linear in the antifields
$\Phi^*_A$ with coefficients $R^A$  equal to the BRS variation
of the corresponding field $\Phi_A$:
\be
  S[\Phi ,\Phi^*] = S_0[\Phi] + \Phi^*_A R^A[\Phi] +
 \ldots,
 \label{ExtAct}
\ee
and additional terms, at least quadratic in the antifields, which encode
special features of the BRST algebra, e.g.\ its off-shell structure.
This action is
to be constructed in such a way that the classical master equation
\be
   (S,S)=0,
\label{CME}
\ee
is satisfied. A solution of this equation
(together with some suitable boundary conditions) always exists
\ct{BV85}, whereas its local character, universally taken as one
of the cornerstones of quantum field theory and therefore a desirable
characteristic, can always be ensured under reasonable conditions on the
gauge algebra generators \ct{Hen91} (see also \cite{ST} and references
therein).

The antibracket, \vgl{antibracket}, the extended action, \vgl{ExtAct},
and the classical master equation, \vgl{CME},
play a central role in the theory. For example,
the antibracket is used to perform canonical transformations of field
variables and to implement gauge fixing (which becomes merely
a change of basis). The extended action is used to
define the BRS operation, and the master equation \bref{CME}
guarantees then its on-shell nilpotence.

\subsection{The quantum master equation and anomalies}

\hspace{\parindent}%
In the quantum theory, equation \bref{CME} is modified. Indeed, from a
purely formal point of view, gauge (or BRS) invariance at quantum level,
with a given action $S$,
is transcribed into the BV formalism as
\be
  \frac{1}{2}(S,S)=i \hbar \Delta S,
\label{formal QME S}
\ee
with
\be
\Delta=\frac{\dl }{\delta\Phi^A}\frac{\dl }{\delta \Phi^*_A}\ .
\label{Delta formal}
\ee
We call attention once again to the implicit notation
(see the footnote on page~\pageref{deWitt-conventie})
which implies that the sum over  $A$ in
\bref{Delta formal} includes not only a
summation over all (components of) the fields, but also an integration over
the space-time point. This point is the same for both functional
derivatives. Accordingly, the expression \vgl{Delta formal} is meaningless
when acting on a local functional --in particular its use in
\vgl{formal QME S}! In fact, it could be viewed as an expression involving
$\delta(0)$, and one may be tempted to neglect it for that
reason, for example appealing to dimensional regularisation \cite{BV}.
However, careful consideration leads one to conclude that in fact
this term is potentially a source of anomalies \cite{anombv}. In the
Pauli-Villars scheme for example, to order $\hbar$,
a definite expression can be associated to it. Then one has to decide
whether this anomaly is genuine: if it can be avoided by adding
an extra counterterm $\hbar M_1$,
the classical action $S$ should
be modified to a 'quantum action' $S_q=S+\hbar M_1+\ldots$
Without restrictions on $M_1$ this is in fact
always possible (see, for instance, \cite{BV}), but here the locality
requirement comes into play again:
for local quantum field theory one will refuse
to add anything but a {\em local} $M_1$.
If one can not find a local $M_1$, the theory has a genuine anomaly,
and \bref{formal QME S} can not be satisfied.
This reasoning can be extended to
higher orders in $\hbar$. The formal quantum master equation is then
\be
\frac{1}{2}(S_q,S_q)=i \hbar \Delta S_q,
\label{formal QME W}
\ee
where $S_q$ is the previously mentioned quantum action%
\footnote{In the sequel, in the quantum context,
we will use simply $S$ instead of $S_q$ to denote the full quantum action.},
formally reducing to the classical action for $\hbar\rightarrow 0$.

In conclusion, within this framework it is the
task of one's regularisation scheme to provide a
well defined replacement for the purely formal equation
\bref{formal QME W}.

\subsection{The Zinn-Justin equation\label{Zinn-Justin subsection}}

\hspace{\parindent}%
In the quantum theory, the master equation can also
be discussed at the level of the effective action
$\Gamma[\varphi,\Phi^*]$, a functional
of the classical fields $\varphi^A$ and of the sources for
the BRST transformations $\Phi^*_A$.
This functional is obtained from the generating functional of the
connected diagrams by trading
the sources $J_A$ for the classical fields $\varphi^A$
by a Legendre transformation, while keeping
the BRST sources $\Phi^*_A$ fixed.  Since $\Gamma$ incorporates
all quantum corrections to the classical action into effective
interactions, i.e. $\Gamma[\varphi,\Phi^*] = S(\varphi,\Phi^*)+
\hbar \Gamma_1[\varphi,\Phi^*] + \ldots$, the quantum BRST
structure is naturally described by the equation
\be
    {1\over 2} ( \Gamma,\Gamma) =
    \frac{\dr \Gamma}{\delta\varphi^A}\frac{\dl \Gamma}{\delta \Phi^*_A}
    = 0 \, ,
\label{QME G}
\ee
of which the classical master equation can be considered to be the
$\hbar = 0$ limit.

Fulfilment of \bref{QME G} ensures in fact gauge independence of the
quantum theory. Instead, if the BRST symmetry is broken by loop
corrections, i.e. the gauge symmetry is {\it anomalous}, equation
\bref{QME G} is replaced, according to Lam's quantum action principle
\ct{lam, Lam2}, by \cite{hlw90}
\be
    {1\over 2} ( \Gamma,\Gamma) = -i \hbar (\cA \cdot\Gamma).
\label{AZJeq}
\ee
On the right hand side, $(\cA \cdot\Gamma)$ is the generating
functional for 1PI diagrams with one insertion of a local
operator $\cA[\varphi,\Phi^*]$. The  BV formalism gives an explicit
albeit formal expression for it,
related to the quantum master equation
\bref{formal QME W}, namely $\cA= \Delta S +(i/2\hbar)(S,S)$. In this
way the operator $\cA[\varphi,\Phi^*]$ makes explicit the breaking of the
classical BRST structure due to quantum corrections and is interpreted as
the BRST anomaly.
The equation \bref{QME G},
in a slightly different guise without the sources
corresponding to what are presently the antifields of the antighosts,
was discovered by Zinn-Justin in a discussion on the
renormalisation of non-abelian gauge theories. It is therefore called the
{\it Zinn-Justin equation} and can be considered as
a rewriting of the Veltman-Ward identity for the BRST symmetry.

A detailed discussion of the BV formalism and
renormalisation using the effective action has been given in
\cite{Anselmi}. Further advocacy for the use of BV in the formulation
of renormalisability can be found in \cite{Gomis-Weinberg}.

\section{BPHZ}
\label{BPHZ}

\hspace{\parindent}%
To put the quantum theory of the BV method on a more solid footing,
one wants a proper quantum version of the basic equations. In particular,
because of its central importance,
we are looking for a regularised version of the quantum master equation
\vgl{formal QME W}
as an operator equation which is valid to all orders in perturbation
theory. In this section we  first discuss some alternatives, and then
give our reasons for using the setup of Bogoliubov, Parasiuk, Hepp and
Zimmerman. We  continue with a summary of the basics of that method,
and recall the treatment of the chiral anomaly as a paradigm.

\subsection{Alternatives}

\hspace{\parindent}%
As was already pointed out, the quantum modifications in
\vgl{formal QME W} as compared to \vgl{CME}, and in particular
a non-zero $\Delta S$, signal the possible
appearance of anomalies. In fact, $\Delta S$ is closely related to the
Jacobian of the path integral measure under BRST transformations,
with a potentially diverging determinant. Let us
therefore consider the method of Fujikawa \cite{fuji}.
The BRST Jacobian determinant is a product of eigenvalues
of some operator in an infinite dimensional space, and this is
regularised by introducing a Gaussian damping, removing the
contribution of the 'high-energy' modes.
However, the method does not provide an unambiguous rule
for choosing the regulator (the Gaussian damping), although
the ad hoc conditions that are sometimes imposed
to obtain a consistent anomaly can be replaced by
a more systematic approach using Pauli-Villars methods \cite{Diaz}.
Also, the extension of the method to higher loops is not straightforward.

In contrast with the philosophy of the Fujikawa method, which
considers the relevant Jacobian determinant as the primary object
to regularise, a more comprehensive point of view is
that a regularisation scheme should take care of the divergences of
the full theory (up to a certain order in perturbation theory).
Most of the universally popular regularisation schemes are formulated as
prescriptions to associate first a regularised, and in a second step a
renormalised, finite expression to diverging loop diagrams.
Let us then consider 'dimensional regularisation', which gives
a method to replace the Feynman rules by an '$n$-dimensional'
generalisation, with $n$ complex. Let us note that
in this scheme one does not have at one's disposal a Lagrangian
corresponding to the regularised step.
It is well known that, when symmetry properties of
the  Lagrangian one considers change with dimension,
dimensional regularisation requires a delicate treatment to provide the
same answers as other regularisation methods, notably when computing
anomalies. In fact, in the initial  BV-treatment
dimensional regularisation was invoked to drop the anomalous terms.
The lack of a Lagrangian seems to prohibit the formulation of
the basic local BV equations in this scheme, although if one accepts
the customary interplay of integer and complex dimensions one should refer
to the treatment%
\footnote{especially charming through its use of $d-4$ as an extra
(global) variable of the theory, with a corresponding BV-antivariable that
plays a role in the treatment of the anomaly too!} in \cite{Tonin}.

We now turn to schemes that {\em do} have a straightforward Lagrangian
interpretation. Originally
due to Pauli and Villars, the simple device of subtracting from loops a
compensating loop in which a particle with a very large mass  circulates
is  the oldest among these. One
writes down a Lagrangian for the original theory {\em plus} an
additional piece where all terms involve one or more new {\em massive}
(Pauli-Villars) fields.
This is done in such a way that at this stage
the theory is (one-loop) finite. Then one computes the effective
action, and does the renormalisation. For a renormalisable theory this
step absorbs the terms that would potentially diverge in the next
step, where one lets all masses of the PV-fields tend to infinity.
The contents of the original theory is defined to be what one obtains in
this limit%.
\footnote{The $\delta(0)$ problem of \vgl{formal QME S} does not arise,
because the offending right hand side can be avoided completely \cite{Frank}
due to a cancellation between the original fields and the PV fields.}.
In this treatment, possible complications arise if the mass terms
do not respect all symmetries of the original theory. Then the classical
master equation is broken
and one has to investigate carefully whether
the breaking terms survive after taking the infinite mass limit. If they
do, the theory may have an anomaly.
This method goes together well with the BV formalism, and
this way of computing the one loop anomaly
is by now well-documented in the literature.
It fits well because the method is based on
a Lagrangian, just like the BV setup%
\footnote{We do not wish to distinguish here between Lagrangian and
Hamiltonian methods, but rather set apart these methods from a purely
diagrammatic approach. We will continue to use the term 'Lagrangian' to
indicate both variants of action-based treatments.}.

The Pauli-Villars scheme only works well at one loop. One could imagine
circumventing this restriction by using higher (covariant) derivatives
in combination with the one loop PV scheme \cite{FadeevSlavnov}%
\footnote{The prescriptions given in \cite{FadeevSlavnov}
are not quite correct, but have
been amended in the recent literature,
see \cite{RuizPietClausBakeyevSlavnov}.}.
However, constructing such covariant derivatives requires an understanding
of the geometry behind the gauge symmetry. For some models, for instance
the $W_3$ gravity model which we use as an example in this paper,
such knowledge may be lacking.

Another resolution to the divergence problem,
operationally very close to Schwinger's proper time method, consists in
modifying the propagators to cut off the loop momentum integrals. The
Lagrangian gauge invariant implementation of this idea is necessarily
nonlocal. This method has been advocated \cite{moffat} as a universal
solution to the regularisation problems, and was
applied to the present problem in \cite{j95}. The most striking virtue of
this method, namely regularisation to {\em all} loop levels, brings also
the computation of higher loop corrections within reach \cite{jw95}, and
makes it into the primary alternative to the method applied in this paper.
We refer to \cite{j95,jw95} for a detailed exposition.

\subsection{Why BPHZ?}

\hspace{\parindent}%
In view of the coherence of the BV formalism, which rest very heavily on
the use of a classical action functional, a properly defined quantum
action functional is highly desirable, and preferably well defined
functional integral expressions as well.
When regularisation is formulated directly in terms of diagrams however,
the path integral is just a shorthand notation for a collection of
diagrams. This notation naturally suggests relations between correlation
functions (=sets of diagrams), such as those obtained by functionally
integrating by parts, the most familiar of these being the
Schwinger-Dyson equations and Ward identities.
Relations of this type are used heavily to set up the formal BV scheme,
so we require that regularisation keeps them in manageable form.

Of all renormalisation methods,
the BPHZ method is probably the most solidly founded one. In fact it is
used as a touchstone (see for example \cite{BreitenlohnerMaison})
for other methods to
establish their respectability. It adresses head-on  the key issue
in the regularisation of quantum field
theory, viz.  the definition of products of local operators at the
same space-time point. In BPHZ this is done via an explicit prescription
of all their matrix elements. These matrix elements are defined in
terms of Feynman diagrams. The diagrams are the same as when
working formally, by which we mean disregarding the divergence problem.
When applying the Feynman rules as they follow from the action,
one obtains meaningless expressions. The BPHZ method is a definite
prescription, order by order in perturbation theory, to supplement the
Feynman rules so that one obtains meaningful expressions. Whereas the
prescription is {\em inspired} by formal manipulations,
'subtracting infinite counterterms', in fact it
makes no reference to such a dubious step. Let us remind the reader that
the best-founded proofs of renormalizability of various theories rest on
the application of this method. Let us also remind the reader that its
prescription is universal, in that it applies irrespective of the order
in perturbation theory.

The big asset of the BPHZ scheme is that renormalized correlation functions
with insertions of arbitrary composite operators can be defined in the
same way and with the same ease as diagrams related to Green's functions
of elementary fields. The technical tool that does the job is a so-called
'normal product' operator, that we discuss in the next subsection. The
normal products uniquely determine the finite expression associated with
every diagram. The relations between correlation functions containing
composite operators that one formally derives from path integrals, all
have counterparts as relations between normal ordered products, and
therefore as relations between renormalized correlation functions. Thus
while the BPHZ method is {\em not exclusively} based on an action
functional, it does provide well-defined quantum operators that
correspond to classical functionals through the normal product concept,
and it does provide counterparts to the classical equations obeyed by
these functionals. The BPHZ scheme is therefore well-suited to formulate
the basic local equations of the BV formalism.

\subsection{Basics of BPHZ renormalization}\label{BPHZ basics}

\hspace{\parindent}%
Elaborating on the work of Bogoliubov, Parasiuk and Hepp \ct{bph},
Zimmerman \ct{z} set up a renormalization scheme in which renormalized
Green functions with composite operator insertions are obtained in
essentially the same way as renormalized Green functions of elementary
fields. The method%
\footnote{Useful general references, containing a more detailed
exposition are \cite{Piguet and Rouet PR,boek Piguet en Sorella}.},
which bypasses the regularization step,
determines the renormalized counterpart $I_R(p)$ of a divergent loop
integral $I(p)$ associated with a specific {\it one particle irreducible}
Green function by subtracting from its integrand the first few terms in
the Taylor series around vanishing external momenta $p$, i.e.
\be
I(p) = \int \dif k \,\, \cI (p,k) \, {\longrightarrow}\,
I_R(p) = \int \dif k \left( 1 - t^\delta_p \right) \cI (p,k)\, .
\label{I1}
\ee
The minimal value of $\delta$ is the superficial
degree of divergence of the diagram (if the diagram
is not finite by power counting). One-loop diagrams can
unambiguosly be treated with \vgl{I1}, but multiloop diagrams with
overlapping divergences require the use of the {\it forest formula}
\ct{z} to determine the Taylor series to be subtracted
(see section \ref{W3} for an explicit example).

In fact, the BPHZ renormalization scheme allows to define for any local
composite operator $\cO(x)$ a sequence of
{\it normal ordered composite operators}
$$
\norm{a}{\cO(x)} \, ,
$$
with $a$ an integer greater than or equal to the
canonical dimension of the operator $\cO(x)$ (counting $1$ for every
derivative, etc.\ ). A normal product is called
{\it minimally subtracted} (resp. {\it oversubtracted}) when $a$ is equal
to (resp. greater than) its canonical dimension.
For a generic connected Green function of free
field normal products $\norm{a_i}{\cO_i(x_i)}$
$$
  \expect{\prod_i\norm{a_i}{\cO_i(x_i)}}_{c},
$$
the order $\delta$ of the Taylor series in Zimmerman's
prescription \bref{I1} is given in terms of the degrees $a_i$ as
\be
   \delta = n + \sum_i ( a_i - n) - \sum_k d_k   \, ,
\label{rule}
\ee
where $n$ is the space-time dimension and the last term subtracts all the
canonical dimensions $d_k$ of external lines.

The use of oversubtracted operators, while seemingly superfluous, is in
fact one of the virtues of the BPHZ method and a cornerstone in its use
for anomaly computations. Moreover,
it is useful to introduce not one but a variety of oversubtracted normal
products \ct{lam}. Indeed, the degree of oversubtraction of an operator
can be propagated to subdiagrams%
\footnote{We always refer to {\em proper} subdiagrams,
i.e. subdiagrams that are themselves 1PI.}
in different ways, the only constraint
being that the degree of oversubtraction of a subdiagram should be at
least as large as the degree of oversubtraction of {\em its} subdiagrams.
When subdiagrams are always maximally oversubtracted
one speaks of {\em isotropic} normal products
(all lines are treated in the same way), otherwise of
{\em anisotropic} normal products. We do not need the most general
anisotropic ones, but only the Gomes-Lowenstein \cite{GomesLowenstein}
(GL) version,
\be
\norm{d_1+d_2+\alpha}{\{\cP_1\}\cP_2} \,,
\label{ani np}
\ee
for which the rule that determines the subtraction degrees
$\delta_\gamma$ of subdiagrams $\gamma$ is modified to%
\footnote{The Taylor series operator acts on all momentum factors
produced by the $\cO_i$, but not on the propagators of the lines
that are external to the subdiagram $\gamma$.}
\be
   \delta_\gamma = n - \sum_k d_k + \sum_{i\neq i_0} ( a_i - n) +
   \left\{
   \begin{array}{ll}
     d_1+d_2 -n, &
     \mbox{if $i_0\in \gamma$ and $L(\cP_2)\subseteq E(\gamma)$} \\
     d_1+d_2 +\alpha-n,&
     \mbox{if $i_0\in \gamma$ and $L(\cP_2)\subseteq \hspace{-3mm}\slash
     E(\gamma)$} \\
     0, &
     \mbox{if $i_0\in\hspace{-3mm}\slash \gamma$ }
  \end{array} \right. \, ,
\label{ani rule}
\ee
where $i_0$ is the normal product vertex under consideration;
$d_l$ the canonical dimension of the operator $\cP_l$;
$L(\cP_2)$, the set of lines associated with $\cP_2$ and $E(\gamma)$ the
set of external lines of the subdiagram $\gamma$.
Note that this rule implies that if $\cP_1$ is linear in the fields then
there is no distinction between the isotropic and the anisotropic
products.

Let us conclude this summary by indicating two
fundamental and useful properties of the
normal ordered composite operators.
The first of these, the BPHZ renormalized version
of the Schwinger-Dyson equations \cite{GomesLowenstein}, is%
\footnote{Here and in the following, to avoid unnecessarily heavy
notations, we act as if there are no dimensionful parameters in the model,
and all operators in the action are minimally subtracted
$\norm{n}{\,\cdot\,}$-products.
It is rather trivial to overcome this restriction.\label{nonminfn}}
\bea
   &\displaystyle{
   i \hbar \expect{ \norm{a}{\cO(x)} \frac{\dl}{\delta \Phi^B(x)}
   (-1)^{\gras{B}}
   \left( \Phi^{A_1}(x_1) \ldots \Phi^{A_m}(x_m) \right)}=}&
\nonumber\\
   &\displaystyle{
   \expect{\left(\norm{a+n-d_B}{ \cO(x)\frac{\dr S_{0}} {\delta \Phi^B(x)}
   + \{\cO(x)\}\frac{\dr \tilde S_{\rm I}} {\delta \Phi^B(x)} }\right)
     \Phi^{A_1}(x_1) \ldots \Phi^{A_m}(x_m)} }&,
\label{ani SD-BPHZ}
\eea
where $d_B$ is the engineering dimension of the field $\Phi^B(x)$;
$S_0$, the massless free part of the action   and
$\tilde S_{\rm I}$ lumps together eventual mass terms and interactions,
i.e. $\tilde S_{\rm I}= S_{\rm I}+ S_m$. Notice the anisotropic normal
ordering in the second term on the rhs.

The second property, the so-called {\it Zimmerman identity},
establishes linear relations among
different normal products of a given monomial.
Different formulations may be given.
A first version relates
normal products with different subtraction degrees
$\alpha_1>\alpha_2$ to all operators $\cO_i(x)$ having the same
quantum numbers and with canonical dimensions $\leq \alpha_1$
\be
   \norm{\alpha_1}{\cO(x)} - \norm{\alpha_2}{\cO(x)} =
   \sum_i r_i \, \norm{\alpha_1}{\cO_i(x)},
\label{ZimId2}
\ee
where the coefficients $r_i$ vanish at lowest order.
Our approach in fact uses a consequence of \bref{ZimId2},
relating oversubtracted operators to their minimally subtracted
counterpart
\be
   \norm{a+\alpha}{\cO(x)} - \norm{a}{\cO(x)} =
   \sum_i r_i \, \norm{a_i}{\cO_i(x)},
\label{ZimId}
\ee
where now {\it all} the operators on the rhs are {\it minimally}
subtracted. A second, extended version of the Zimmerman identity
further relates anisotropic and isotropic normal products with the
{\it same} substraction degree
\be
    \norm{d_1+d_2+\alpha}{\{\cP_1\}\cP_2}-
    \norm{d_1+d_2+\alpha}{\cP_1\, \cP_2}=
    \sum_i r_i \norm{a_i+d_2}{\cO_i(x)\cP_2},
\label{aniZim}
\ee
where the same restrictions for $a_i$ and $r_i$ apply also in this case.
In the next section, the two relations \bref{ani SD-BPHZ} and
\bref{ZimId} will be used extensively for the implementation
of BPHZ ideas into the BV framework.

\subsection{Anomalies and the massless case}
\label{Aandmass}

\hspace{\parindent}%
The exposure of the chiral anomaly in the present framework is an
application of the BPHZ method that dates back some 25 years. We shall
not repeat this calculation here, since a short account of it can be found
in \cite{bron}, but just sketch the main ideas and stress that such recipe
works in general for determining the anomaly in a global, continuous
symmetry.

The central equation, which computes the divergence of (matrix elements
of) the axial vector current in Q.E.D.
(\cite{bron}, eqs.\,(1.7) and (1.8)) is
\bea
   &\displaystyle{
\partial_\mu \expect{\norm{3}{\bar\psi \gamma^\mu\gamma^5 \psi}(x) X}=}&
\nonumber\\
   &\displaystyle{
  -\expect{\left[(\gamma_5\,\psi(x)) \frac{\delta}{\delta\psi(x)}
   +(\bar\psi(x)\,\gamma_5) \frac{\delta}{\delta\bar\psi(x)}\right] X}
   +2im \expect{\norm{4}{\bar\psi \gamma^5 \psi}(x) X}, }&
\label{Low1.7}
\eea
which is an application of \vgl{ani SD-BPHZ}.
Then the oversubtracted operator
$\norm{4}{\bar\psi \gamma^5 \psi}$ is rewritten in terms of minimally
subtracted ones by using Zimmerman's identity \bref{ZimId} as
\be
 \norm{4}{\bar\psi \gamma^5 \psi}=
 \norm{3}{\bar\psi \gamma^5 \psi}+
 r\,\norm{4}{F_{\mu\nu}\tilde F^{\mu\nu}}+
 s\,\norm{4}{\partial_\mu(\bar\psi \gamma^\mu\gamma^5 \psi)}.
\label{Low1.8}
\ee

An important feature of the treatment is that,  for massless fermions,
a mass has to be introduced to perform the calculation. The reason is
that the BPHZ method introduces an apparent infrared divergence in the
subtraction terms, in spite of the fact that no physical infrared
problems are associated with the vanishing of the electron mass.
It comes up when one tries to calculate the
matrix elements of the oversubtracted operator
$\norm{4}{\bar\psi \gamma^5 \psi}$: they diverge
in the massless limit. Ultimately, this is
why one is {\em forced} to reduce the oversubtracted to an
ordinary normal product: the latter has matrix elements that stay finite
when the mass goes to zero
(if there is no physical infrared problem).
The upshot in the present computation is that, in \vgl{Low1.7},
when $m\rightarrow 0$, one picks up the terms from \vgl{Low1.8} that
go as $m^{-2}$. This is the origin of the axial anomaly.
For the actual computation we refer to the quoted literature for
the axial anomaly, and for other examples
to sections~\ref{1law2mt}~and~\ref{2law3mt} and appendix~\ref{ghanom}
of this paper.

\section{Anomalies in the BV--BPHZ renormalized framework}
\label{BVandBPHZ}

\hspace{\parindent}%
The properties of the normal ordered products \bref{ani SD-BPHZ},
\bref{ZimId2} have
been known to provide a natural framework for the
investigation of anomalies in global symmetries for a long time.
Furthermore, they were extensively used in the original presentation
of the BRS quantization scheme \ct{brs}.
We will carry these ideas one step further by implementing BPHZ
ideas in the BV framework, in order to formulate its basic local equations
--essentially the quantum master equation-- on the level of renormalised
perturbation theory and to derive a proper method to compute higher loop
anomalies. For treatments on the level of the effective action, which have
been limited to one-loop anomalies, we refer
to~\cite{boek Piguet en Sorella} and references therein.

Roughly speaking, this program is realized by
substituting formal products of fields by suitable Zimmerman normal
products. In the assignment of dimensions $d^A$, $d^*_A$ to the fields
$\Phi^A$ and antifields $\Phi^*_A$, there is a certain degree of
arbitrariness, although they are largely fixed by the classical free
Lagrangian. This arbitrariness is fixed in a way that is convenient for
the BV framework by
demanding that the BRST operator preserves canonical dimensions%
\footnote{If one follows the traditional assignment,
of \cite{bron} for instance, the usual Yang-Mills ghosts and antighosts
would both have (scaling) dimension $1$, whereas we assign dimension $0$
to $c$ and $2$ to $b$, as in ref.\,\ct{boek Piguet en Sorella}. Also,
antifields would have $0$ dimension (being external fields), whereas our
antifields do not. The BRST operator increases that (traditional) scaling
dimension by $1$ unit. We will adopt conventions matching those
of conformal field theory (see
table~\ref{tbl:charges} on page~\pageref{tbl:charges} and
table~\ref{tbl:chargesII} on page~\pageref{tbl:chargesII}).
The difference is only one of bookkeeping.}.
The following relations then hold
\be
        d\left[ \frac{\delta S}{\delta\Phi^A} \right] = n - d^A,
        \quad\quad
        d\left[ \frac{\delta S}{\delta\Phi^*_A} \right] = d^A=n -d^*_A\, ,
\label{dimensions}
\ee
such  that the dimensions of a field and of its corresponding antifield
add up to the space-time dimension $n$.
As a consequence, normal products involving BRST sources obey, for
example
\be
    \norm{a}{\Phi^*_A R^A}=\Phi^*_A \norm{a-n+d_A}{R^A}.
\label{rel anti}
\ee

An essential ingredient to deal with massless theories within our program
consists in the introduction of a small mass $m$ for the massless fields.
As was pointed out in sec.\,\ref{Aandmass}, this is necessary to
prevent the appearance of spurious IR divergences when performing the
prescribed subtractions at vanishing external momenta%
\footnote{Subtraction at non-zero momentum is feasable, but unnecessarily
complicated regarding counterterms and Lorentz invariance.}.
In these cases, one therefore
replaces the original massless action $S$ by a massive action $\tilde S$
\be
  \tilde S(\Phi,\Phi^*)=S(\Phi,\Phi^*)+{m^2\over 2} S_m(\Phi)=
  S + {m^2 \over 2} \Phi^A T_{AB} \Phi^B,
\label{massive action}
\ee
supplemented with the rule that the limit $m \rightarrow 0$ is taken at the
very end of the renormalized computations.
Also, $d^A + d^B + d(T_{AB}) = n-2$.

We will now formulate the basic
equations governing the BV method at the renormalised perturbative level.
As already pointed out in the introduction, in the BV scheme
the quantum BRST structure and its possible violations
are naturally described by means of the BRST-Veltman-Ward
identity \bref{AZJeq} for the effective action $\Gamma$.
To write down the corresponding local equation
we start from the
generating functional of {\em all} connected Green functions $W(J)$,
of which $\Gamma$ is the Legendre transform,
$\Gamma(\Phi,\Phi^*) = W(J,\Phi^*) -J \Phi$,
where it is understood that to obtain the left hand side one should solve
the equation $\Phi=\frac{\delta W}{\delta J}$ for $J$ as a function of
$\Phi$ and $\Phi^*$. By inverting the Legendre transform, one then also
has that $\frac{\delta \Gamma}{\delta \Phi}=-J$. The left hand side of
equation \bref{AZJeq} is then equal to
\be
    -\int \dif^n x \,\,
    J_A(x) \frac{\dl \Gamma}{\delta\Phi^*_A(x)}.
\label{int0}
\ee
A general property of Legendre transformations is that the derivatives of
$\Gamma$ and $W$ with respect to any parameter, $\Phi^*$ for example,
are equal:
\be
\frac{\dl \Gamma(\Phi,\Phi^*)}{\delta\Phi^*_A(x)}=
\frac{\dl W(J, \Phi^*)}{\delta\Phi^*_A(x)},
\ee
where on the l.h.s $\Phi$ is kept constant in taking the derivative, and
on the r.h.s. $J$ is kept constant, and it is understood that after taking
the derivatives the functional relation between $J$ and $\Phi$ is imposed.
Since $W$ is nothing but the generating functional of all the connected
Green functions, a (formal) power series in the source $J$, its derivative
with respect to the parameters $\Phi^*$ is the generating functional of
all connected Green functions with one extra insertion of the operator
$\frac{\dl S}{\delta\Phi^*_A(x)}$.
Suppressing the $\Phi^*$ dependence in the notation, we will denote
the resulting generating functional
with the insertion of $X$ by $\expect{X}_{c,J}$,
so that, for example, $W(J,\Phi^*)=\expect{\bf 1}_{c,J}$.
This local functional of the
fields (and possibly of antifields as well) that one has to insert
is not just the classical functional (which would be ill-defined),
but one of the normal products that correspond to this classical
functional, which have a definite meaning in the BPHZ scheme.
The particular normal product one takes is an integral part of the
definition of the quantum action functional one starts from.
We take the interaction terms in $S$, in particular the antifield terms,
to be minimally subtracted.
Other choices differ in terms of order $\hbar$ or higher,
see~\bref{ZimId}. One may wish to include
in the action more general terms of higher order in $\hbar$.
This would be the case when loop anomalies are to be canceled by finite
non-invariant counterterms (usually called $M_1, M_2, \ldots$ in quantum
BV formalism). We will continue to denote the full quantum action,
including these terms if they are introduced, by the symbol $S$.
(If some of these terms contain antifields, they are to  be viewed
as quantum modifications of the BRST transformation laws.)
The normal ordering to be used for the
lhs is then, according to \bref{dimensions}, $\norm{d^A}{\cdot}$.
The equation  \bref{AZJeq} is then completely equivalent with
\begin{equation}
    \int \dif^n x \,\, J_A(x)
    \expect{ \norm{d^A}{\frac{\dl S}{\delta\Phi^*_A(x)}}}_{c,J} =
    i \hbar \int \dif^n x \,\, \langle \norm{n}{\cA (x)}
\rangle_{c,J} \, .
\label{intx}
\end{equation}
It should be emphasized that this equation is an exact one, involving on
the l.h.s. only quantities which are well-defined in renormalised
perturbation theory, although rather empty until one specifies the
operator $\norm{n}{\cA (x)}$ on the r.h.s.: in fact one may take it as
the definition of $\cA$.
If one treats the equation {\em formally}, ignoring the normal product
rules, for example using the heuristic
path integral formulas, then, as already mentioned in
section~\ref{Zinn-Justin subsection},
it is rather easy to derive the general
formal expression $\cA=\Delta S +\frac{i}{2\hbar}(S,S)$, for which
there is of course no place in the BPHZ scheme.
Reverting to a more proper treatment, there is still a general statement
that can be made:
a celebrated result of general BPHZ renormalisation theory, namely
Lam's theorem \cite{lam, Lam2}
(also called the quantum action principle) guarantees that
the anomaly insertion $\cA(x)$ in \bref{AZJeq}
is the integral of a {\it local}
composite operator, whose canonical dimension
is  $n$ in our conventions%
\footnote{The well-known $\gamma_5$ anomaly in $4$ dimensions,
including the factor $c$, would have canonical dimension $5$
if we had adopted the counting with the usual scaling dimensions.}%
, without specifying its explicit form.
We will not need to rely on this general theorem here,
since we will now deduce an explicit formula for $\cA$
that shows this locality explicitly.

Expanding \bref{intx} in the sources $J_A$ one obtains
\bea
    &\displaystyle{
    - \int\dif^n x
    \expect{ \left(\norm{d^A}{\frac{\dl S}{\delta\Phi^*_A(x)}}
    \frac{\dl}{\delta\Phi^A(x)} (-1)^{\gras{A}}\right)
    \Phi^{A_1}(x_1) \ldots \Phi^{A_m}(x_m)}_c}&
\nonumber\\
    &\displaystyle{
    =\int \dif^n x \,\,
    \expect{\norm{n}{\cA (x)}
    \Phi^{A_1}(x_1) \ldots \Phi^{A_m}(x_m)}_c}.&
\label{anomeq1}
\eea
These equations are nothing but the typical anomalous Ward identities
for the usual BV-BRST transformation $\delta$ since, for antifield
independent functionals, one has classically that
\be
   \delta F[\Phi]=
    \int\dif^n x \left\{\frac{\dl S}{\delta\Phi^*_A(x)}
    \frac{\dl}{\delta\Phi^A(x)} (-1)^{\gras{A}}\right\}F[\Phi]=
    (F,S).
\label{brst op}
\ee

To obtain from \vgl{anomeq1} an equation in terms of local quantum
operators (instead of correlation functions), we
apply the BPHZ expression
of the Schwinger-Dyson (SD) equation, in the form given
by Gomez and L\"owenstein, \bref{ani SD-BPHZ}, to its lhs,
for the operators $\gh O(x)=\frac{\dl S}{\delta\Phi^*_A(x)}$.
This leads to the following operator identity in terms of
(possibly oversubtracted) anisotropic normal products
\bea
   &\displaystyle{
   - i \hbar \int \dif^n x \,\, \norm{n}{\cA (x)}=}&
\nonumber\\
   &\displaystyle{ \int \dif^n x \,\, \left(
   \norm{n}{\frac{\dr S_0}{\delta\Phi^A(x)}
   \frac{\dl S}{\delta\Phi^*_A(x)}}+
   \norm{n}{\frac{\dr (S_m + S_I)}{\delta\Phi^A(x)}
   \left\{\frac{\dl S}{\delta\Phi^*_A(x)}\right\} }\right)
    \,}& .
\nonumber
\eea
Assuming the validity of the classical master equation
\bref{CME} for $S= S_0 + S_I$, the functionals inside the
normal ordering can be simplified, resulting in
\bea
   &\displaystyle{
   - i \hbar \int \dif^n x \,\, \norm{n}{\cA (x)}=}&
\nonumber\\
   &\displaystyle{ \int \dif^n x \,\, \left(
   \norm{n}{\frac{\dr S_I}{\delta\Phi^A(x)}
   \left\{\frac{\dl S}{\delta\Phi^*_A(x)}\right\} } -
   \norm{n}{\frac{\dr S_I}{\delta\Phi^A(x)}\frac{\dl S}
   {\delta\Phi^*_A(x)}} \right) }& \nonumber \\
   &\displaystyle{  + \int \dif^n x \,\,
   \norm{n}{\frac{\dr S_m}{\delta\Phi^A(x)}
   \left\{\frac{\dl S}{\delta\Phi^*_A(x)}\right\} }. }&
\label{anomeq2}
\eea

This is the  promised explicit local expression for the
BRST anomaly in the BPHZ framework.
We now comment on the different contributions in \bref{anomeq2}.
The first contribution,
potentially present in any interacting theory, comes from
the difference between the anisotropic and the isotropic normal products of
the BRST variation of the interaction term $S_I$, as a straightforward
application of the generalized Zimmerman identity \bref{aniZim} shows.
These contributions vanish however for {\it minimally} subtracted%
\footnote{In \cite{GomesLowenstein} it is indicated that in nonlinear $\sigma$-models
one may prefer non-minimal interaction terms. Therefore, these terms can not
be dropped in general, although in the examples of this paper they
could. See also footnote~\ref{nonminfn} on page~\pageref{nonminfn} }
BRST variations of the interaction, since then the GL
anisotropic normal products (\ref{ani np},\ref{ani rule}) and the isotropic
ones exactly coincide.
The second  contribution to the anomaly
originates in the explicit breaking of the BRST (or gauge)
symmetry by the IR regulating mass term $S_m$ in
\bref{massive action}
\be
    \lim_{m\rightarrow 0}\,
     m^2 \int \dif^n x \,\, \norm{n}{\Phi^A T_{AB}
     \left\{\frac{\dl S}{\delta\Phi^*_B }\right\} (x)} \, .
\label{nint2}
\ee
The massless limit may result in a non-zero contribution,
as can be seen through a generalization of the argument
presented in sec.\,\ref{Aandmass}.
The composite operator in \bref{nint2} has dimension
$n-2$ (or $n-1$ for Fermi fields, for which the typical mass term in
\bref{massive action} would be proportional to $m$ instead
of $m^2$). Due to the normal ordering degree $n$,
its insertion in correlation functions
leads to {\it oversubtracted} integrals, so that taking the limit
$m\rightarrow 0$ naively by simply discarding such contributions is
incorrect. To compute this limit correctly requires a conversion
to minimally subtracted operators first. We do this conversion in
two steps. First we use  the Zimmerman identity \bref{aniZim} to relate
the anisotropic, oversubtracted normal product to its isotropic,
oversubtracted counterpart, where the normal product corrections
are expressed in terms of operators $\cM_k(x)$ with
canonical dimension $d_k\leq n$, i.e.
$$
   \norm{n}{\Phi^A T_{AB} \left\{\frac{\dl S}{\delta\Phi^*_B }\right\}(x)}
   = \norm{n}{\Phi^A T_{AB} \frac{\dl S}{\delta\Phi^*_B}(x)}
   + \sum_k b'_k\, \norm{d_k}{\cM_k(x)} \, .
$$
In a second step, the isotropic, oversubtracted operator is expressed in
terms of minimally subtracted normal products by means of the
second Zimmerman identity \bref{ZimId}. The net result of this process is
then written as
\be
   \norm{n}{\Phi^A T_{AB} \left\{\frac{\dl S}{\delta\Phi^*_B }\right\}(x)}
   = \norm{n-2}{\Phi^A T_{AB} \frac{\dl S}{\delta\Phi^*_B}(x)}
   + \sum_k b_k\, \norm{d_k}{\cM_k(x)},
\label{Zimmermanid}
\ee
where the numerical coefficients $b_k$ are at least of first
order both in coupling constants and in $\hbar$.
After substitution of \bref{Zimmermanid} in \bref{nint2},
the limit $m\rightarrow 0$ can be taken.
If the theory has no infrared singularities
the minimally subtracted
normal products have finite matrix elements and consequently
only those terms will contribute to \vgl{nint2} that have
coefficients $b_k$ of order $1/m^2$.
The corresponding operators determine the second contribution to the
anomaly in \vgl{anomeq2}.
If there are infrared singularities, the effective action
$\Gamma$ strictly speaking does not exist. A treatment of a
Wilson-type effective action can certainly be developed,
and a corresponding local operator treatment, and it seems reasonable
to anticipate that, as long as the infrared divergences are merely
logarithmic the treatment given above remains valid

To conclude our theoretical approach, we remark that the contribution to
the anomaly generated by the IR regulating mass term is not uniquely
fixed. Indeed, there may be more than one way to construct a mass term in
\bref{massive action}. Different mass terms may behave differently under
some symmetries, causing the anomaly to show up in different
symmetries. A similar ambiguity is present in the context of the
one-loop Pauli-Villars regularisation (see \ct{Diaz, anombv},
and \ct{interpol} for an example how this can be exploited),
but we have not investigated whether
this correspondence is complete in all details.

In summary, the simple rule of substituting formal products of fields by
suitable Zimmerman normal products has led us to an explicit
expression, \vgl{anomeq2}, for the anomaly in the
BV-BPHZ renormalized framework, which is theoretically quite
interesting but not necessarily practical. In fact, it turns out that
for the actual computation of the BRST anomaly insertion $\gh A$
in \bref{AZJeq}, \vgl{anomeq1} is more convenient once an
enumeration of possible monomials in $\gh A$ has been given.
The next section will illustrate this by computing the
conformal anomaly in chiral $W_2$ gravity using both alternatives.
This will of course also serve as an illustration for the more
abstract development of this section. In the section after that, we will go
on and apply the method to the one and two-loop anomaly of $W_3$ gravity,
mainly using the anomalous Ward identities \bref{anomeq1},
but the two-loop anomaly will also be derived from \vgl{anomeq2}.

\section{The anomaly of chiral $W_2$ gravity}
\label{W2}

\subsection{The model}

\hspace{\parindent}%
Chiral $W_2$-gravity is a $2$--dimensional model $(n=2)$
describing $D$ matter fields $\phi^i$, $i=1,\ldots,D$, coupled to a
gravitational gauge field through their energy-momentum tensor
$T=1/2 (\partial\phi^i)(\partial\phi^i)$.
A convenient, gauge-fixed extended action%
\footnote{We take this at the same time to be the {\em quantum} action,
and as in the general treatment of section~\ref{BVandBPHZ}
the minimally subtracted normal products are understood.
With this understanding, and including no modifications $M_i$ of
higher order in $\hbar$, it will turn out that the anomaly comes
out in its conventional form.\label{S en Sq}} for this model can be
taken from \cite{ST}~:
\bea
   S=\int\dif^2 x\hspace{-5mm}&&\left\{
   \left[-\frac12(\partial\phi^i)(\bar\partial\phi^i)
   +b(\bar\partial c)\right]\right.
\nonumber\\
   && \left.
   +\phi^*_i\left[c(\partial\phi^i) \right]
   +b^*\left[-T +2 b(\partial c)+ (\partial b) c \right]
   +c^*\left[ (\partial c) c\right]\right\}
\nonumber\\
     &=& S_0+\int\dif^2 x\,\, \Phi^*_A(x) R^A(x) \, .
\label{W2action}
\eea
In writing \bref{W2action} the following conventions have been used:
$$
  \partial=\partial_+,\quad\quad \bar\partial=\partial_-,\quad\quad
  x^{\pm}=\frac1{\sqrt2}(x^1\pm x^0) .
$$
The fields  $c$ and $b$ in \bref{W2action} are respectively the ghost and
antighost of the spin $j=2$ gauge symmetry, and
$\{\phi^*_i, c^*, b^*\}$
are the antifields or sources for the BRST transformations for all
fields. Remarkably, in this formulation interactions are completely
contained in the antifield dependent terms, so that BRST sources can be
regarded as coupling constants, in which perturbative
expansions can be performed. In fact, the antifield $b^*$ appears in
\bref{W2action} as the source for (minus) the total energy-momentum tensor
of matter and ghosts, viz.
\be
    R^b = - {1 \over 2} (\partial \phi^i) (\partial \phi^i)
    + 2 b \partial c + (\partial b)c =
    -T -T_{\rm gh}\equiv -T_t \, .
\label{Ttot}
\ee
Thus $b^*$  takes over the r\^ole of (minus) the gravitational
gauge field.

We will work with the more general expression for the
ghost energy momentum tensor
\be
          T_j = -j b \partial c + (1-j) (\partial b) c   \, ,
\label{Teejay}
\ee
for a generic spin $(j, 1-j)$ $bc$-ghost system.
At the moment we need $j=2$; in the next section, for
chiral $W_3$, we will have both spin $j=2$ and spin
$j=3$ ghost sectors.

For the assignment of dimensions, we follow common practice in conformal
field theory. There are two conformal dimensions for each field,
corresponding to the left and right Virasoro algebras. The total dimension
$d$ is the sum of these, the spin $j$ is the difference. The relevant
assignments are collected in table \ref{tbl:charges}.
For the antifields, we extended
the assignments in accordance with \vgl{dimensions}, and spin conservation.
Note that the fields and the $\partial$ derivative have vanishing $d-j$,
whereas antifields and the $\bar\partial$ derivative all have
$d-j=2$. This will be useful in our analysis.

These dimensions do not correspond to the assignments used in
section~\ref{BPHZ basics}:
a canonical assignment would attribute dimension $1/2$ to
all ghost fields. The subtraction rules in
\vgl{rule} and \vgl{ani rule} require modifications that reflect this.
We quote the resulting modification only for the case of operator products
that do not contain antifields (which should be extracted first using
\vgl{rel anti}), both for $W_2$ and $W_3$:
\be
   \delta = n + \sum_i ( a_i - n) +  {\sum_k}' (d_k -1) \, ,
\label{j rule}
\ee
where the primed sum  runs {\em only} over ghost
external lines. This new rule extends to the anisotropic subtraction
degree \bref{ani rule}, in which
$\sum_k d_k$ should be replaced by $\sum'_k (-d_k + 1)$ as well.

%%%%%%%%%%%%%%%%%%%%%%%%%%%%%%%%%%%%%%%%%%%%%%%%%%%%%%%%%%%%%%%%%%%%%%%%%%%
\begin{table}[htf]
\begin{center}
\begin{tabular}{||c||c|c|c|c||}
\hline
 & dim $d$ & spin $j$ & gh.\,$\sharp$ & $d-j$  \\
\hline \hline
$\phi$& 0&0&0&0 \\
$c$ &-1&-1&1&0 \\
$b$ & 2&2&-1&0 \\
\hline
$\phi^*$  & 2 & 0 &-1 & 2 \\
$c^*$ & 3 & 1 & -2 & 2  \\
$b^* $& 0 & -2 & 0 & 2 \\
\hline
$\partial$ & 1 & 1 & 0 & 0 \\
$\bar \partial$ & 1 & -1 & 0 & 2 \\
\hline
    \end{tabular}
\caption{Additive charges of fields and antifields in the $W_2$ model.
\label{tbl:charges}}
\end{center}
\end{table}
%%%%%%%%%%%%%%%%%%%%%%%%%%%%%%%%%%%%%%%%%%%%%%%%%%%%%%%%%%%%%%%%%%%%%%%%%%%

Finally, for the reason mentioned in section~\ref{Aandmass}, we introduce
a mass term for all
propagating fields. For the matter fields, this is conveniently done by
including in $S_m$ \bref{massive action} the term
\be
   - {m^2 \over 2} \int d^2 x \,\, \phi^i(x)\phi^i(x) \, .
\label{mama}
\ee
The free propagator for the massive matter fields then becomes
\be
    \expect{\phi^i(x)\phi^j(y)}_0\equiv
    \Delta^{ij}(x-y) = \delta^{ij}
    { i\hbar \over \partial \bar \partial - m^2} \delta(x-y),
\label{proma}
\ee
with the subscript $0$ indicating from now on {\it free correlation
functions}, that is, the ones evaluated in the zero-th order
in the perturbative expansion in antifields.
Introducing a mass term for the ghosts is slightly more subtle. The
obvious choice, $- m\,\int d^2 x \,\, b c$, has the inconvenience of
breaking spin $(j)$ invariance. This  can  be amended by
using the non-local mass term
\be
   - m^2  \int d^2 x \,\, b (x) {1 \over \partial } c(x) \, .
\label{magh}
\ee
The introduction of this nonlocality requires some comments.
The extra mass terms recall a similar feature in Pauli-Villars
regularisation. There, extra fields are introduced to regularise loop
integrals, and these extra fields are given very large masses at the end
of the calculation. Anomalies are understood as remnants of the fact that
these mass terms do not respect all symmetries that were present
classically. For example, in \cite{Diaz} the ghost number anomaly was
traced to a mass term with ghost number different from $0$.
There are some important differences with the  BPHZ method followed
in this paper.
The most important one is that now the masses are given to the
{\em original} fields (there are no extra fields), and they tend to zero
at the end, not infinity.
Thus  the nonlocality of the mass term does not pose any problem
in the present framework, whereas in the PV scheme it would.
The locality is preserved here by the subtraction method (a Taylor series
in momenta), and the introduction of masses is a device needed when
oversubtracted operators appear. The consistency of the setup can be
verified by checking that, despite having ghost number zero, this mass
term reproduces correctly the well-known ghost number anomaly. This is
demonstrated in appendix \ref{ghanom}. In fact the
peculiar non-locality is at the origin of the ghost number
anomaly. From a perturbative point of view, the nonlocal character
of \bref{magh} poses no problem either.
With this choice, the free propagator for the massive ghost fields
\bea
   &\displaystyle{
   \expect{c(x) b(y)}_0 = \expect{b(x) c(y)}_0 =
   -\expect{b(y) c(x)}_0}&
\nonumber\\
   &\displaystyle{
   \equiv G(x-y) =
   { i\hbar \partial_x \over \partial \bar \partial - m^2} \delta(x-y).}&
\label{progh}
\eea
takes a form that is common for Fermi fields. We conclude
that the ghost mass term \bref{magh}, though non-local, is
indeed appropriate for our purposes.

\subsection{The $W_2$ anomaly from the anomalous Ward identities}
\label{w2awi}

\hspace{\parindent}%
Now everything is ready to start the computation of the anomaly
by using the anomalous Ward identities (\ref{anomeq1}).
Its evaluation can be divided in two steps. First, one uses all possible
information concerning symmetries and quantum numbers to determine the
general form of the monomials that can build up the anomaly.
The connected character of the involved Green functions together with
the presence of the normal products severely restricts the form of the
anomaly candidates, so that their field content is completely determined.
A standard perturbative computation then determines  the coefficients
of such anomaly candidates.

Let us now implement this program in the present example.
Since the BRST transformation $\delta$, \vgl{brst op}, preserves spin
$(j)$ the anomaly is a spin zero object, $j[\cA (x)]=0$.
Combining this information with the canonical
dimension, $d[\cA (x)] = 2$ in our conventions,
one finds $(d-j)[\cA (x)] = 2$.
{}From Table \ref{tbl:charges} it follows that all possible
terms in the anomaly contain {\it either}
precisely one antifield {\it or} precisely one $\bar \partial$-derivative.
Lam's theorem further ensures  that the numerical coefficients
of the candidate anomalies are at
least of first order in both $\hbar$ and the coupling constants, the
antifields $\Phi^*$. Thus the anomaly has no $\bar \partial$ dependent
terms, and is of the form
$\cA (x) = \Phi^*_A(x) F^A(\partial, \Phi^B; x)$. This
determines the antifield dependence of the anomaly completely.

Consider now the anomalous Ward identities \bref{anomeq1}. From the
the above analysis, it is clear that all the relevant
information about the functions $F^A(x)$ --up to now, still arbitrary
functions of $\partial$ and all fields-- is contained in the
loop contributions to the linear term in the antifield expansion
of the lhs of \bref{anomeq1}, namely
\bea
   &\displaystyle{\int \dif^2 y
   \expect{ \norm{d^A}{R^A(x)}
   \left(\norm{d^B}{R^B(y)}
   \frac{\dl}{\delta\Phi^B(y)}(-1)^{\gras{B}}\right)
   \Phi^{A_1}(x_1) \ldots \Phi^{A_m}(x_m)}_{c,0}}&
\nonumber\\
   &\displaystyle{= i\hbar\expect{\norm{d^A}{F^A(x)}
   \Phi^{A_1}(x_1)\ldots \Phi^{A_m}(x_m)}_{c,0}\,,}&
\label{ord1bis}
\eea
where relation \bref{rel anti} was used. The form of the functions
$F^A(x)$ can now be completely determined as follows. On the one hand, the
normal ordered character of the insertions in \bref{ord1bis} and
the connectedness of the Green functions indicate that
``proper'' loops (i.e.\,no tadpoles, which are zero due to the
subtractions) can only be generated by making {\it at least} two
contractions between the two BRST transformations $R^A$ and $R^B$ on the
lhs of (\ref{ord1bis})%
\footnote{As a consequence linear transformations do not contribute.}.
Their quadratic field dependence restricts
moreover the maximum number of contractions to be precisely two and leads
thus to the conclusion that {\it only one-loop anomalies} can appear. On
the other hand, from inspection of the free propagators \bref{proma},
\bref{progh}, it follows that these double contractions are only possible
when both $A$ and $B$ refer to the antighost $b$.
These contractions made, no fields are left to
contract with, such that the ``test product'' of fields
$\Phi^{A_1} \ldots \Phi^{A_m}$ must be equal to $b(z)$ in order to get a
non-vanishing lhs. The lhs of (\ref{ord1bis}) becomes
$$
      - \int \dif^2 z
      \expect{ \norm{2}{R^b(x)}
      \norm{2}{R^b(z)} \frac{\dl}{\delta b(z)}\, b(y)}_{c,0}=
     - \expect{ \norm{2}{T_t(x)} \norm{2}{T_t(y)}}_{c,0}  \, ,
$$
since $R^b= -T -T_{\rm gh}\equiv -T_t$, see \vgl{Ttot}.

\begin{figure}
\epsfbox{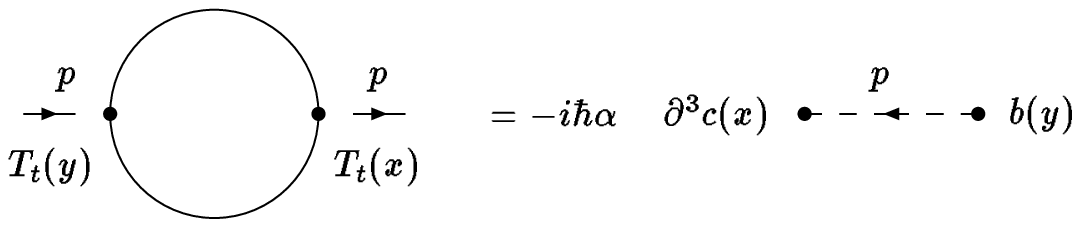}
\caption{The one-loop anomaly diagram for $W_2$.}\label{W2figuur}
\end{figure}

The correlation function on the rhs of
\bref{ord1bis}, of the form $\expect{ \norm{2}{F(x)} b(z)}_{c,0}$ with the
above test product, is non-vanishing if and only if $F(x)$
contains precisely one $c$-ghost, since
the presence of other fields in $F(x)$ would lead to extra
tadpoles. Spin conservation implies
$\gh A(x)= \alpha b^* F(x)=\alpha b^*\partial^3 c $,
so that \bref{ord1bis} becomes
\be
        \expect{ \norm{2}{T_t(x)} \norm{2}{T_t(y)}}_{c,0}
        = - i \hbar \alpha \langle
       \norm{2}{\partial^3 c(x)} b(y) \rangle_{c,0} \, .
\label{TT}
\ee
This is an explicit equation for the anomaly coefficient
$\alpha$, which corresponds diagramatically to Figure~\ref{W2figuur}.
The coefficient $\alpha$ is thus determined by double contractions of the
total energy-momentum tensor $T_t$ with itself. Let us stress that the
lhs of \bref{TT} is completely determined in renormalized perturbation
theory, owing to the normal orderings. The equation expresses in the BPHZ
language the well-known fact that double contractions in the OPE of
$T_t(x)T_t(y)$ generate the conformal anomaly.

The anomaly computation is rounded off with the determination of
the actual value of the coefficient $\alpha$ from eq.\,\bref{TT}.
On the lhs, the matter
and ghost field contributions can be analyzed separately,
since no contractions between $T$ and $T_{\rm gh}$ are possible.
The {\it formal} expression for the contribution
of the matter sector is
\bea
     \langle T(x) T(y) \rangle_{0}  & = & - \hbar^2 {D \over 2}
     \left[
     {\partial^2 \over \partial \bar \partial - m^2} .\delta(x-y)
     \right]^2
\nonumber \\
     & = & - \hbar^2 {D \over 2}
     \int {\dif^2 p  \over (2\pi)^2}  e^{i p\cdot(x-y)} I_{22}(p,m) \, ,
\nonumber
\eea
where $ I_{22}(p,m)$ is a quadratically divergent integral (the
explicit expression is in \vgl{gen1Lint} with $a=b=2$).
The correct expression is similar, and arises
by replacing the composite operators $T$
by their (minimally) normal ordered counterparts,
$\norm{2}{T}$. The subtraction rule
\bref{j rule} states that the integrand $\cI_{22} (q,p,m)$ in
the quadratically divergent one-loop integral
must be replaced  by $(1 -t^2_p) \, \cI_{22} (q,p,m)$,
confirming that minimally subtracted operators yield subtraction
degrees precisely equal to the superficial degree of divergence.
The  correlation function of the normal ordered
energy-momentum tensor for the matter sector is
\be
    \langle \norm{2}{T(x)} \norm{2}{T(y)}\rangle_{c,0} =
           - \hbar^2 {D \over 2} \int {\dif^2 p  \over (2\pi)^2}
           e^{i p\cdot(x-y)}  (1 - t^2_p) \cdot I_{22}(p,m) ,
\label{NOTTm}
\ee
where we introduced the  short hand notation
\be
    I^R_{ab}(p,m)\equiv (1-t^n_p)\cdot I_{ab}(p,m) \equiv
    \int {\dif^2 q \over (2\pi)^2} (1-t^n_p)\, \cI_{ab}(q,p,m),
\label{shn}
\ee
and $\cI_{ab}(q,p,m)$ is the integrand of the formal integral
\be
    I_{ab}(p,m) = \int {\dif^2 q \over (2\pi)^2} { q^a (p-q)^b
    \over (q\bar q+m^2)[(p-q)(\bar p- \bar q) + m^2 ]}\equiv
    \int {\dif^2 q \over (2\pi)^2}\, \cI_{ab}(q,p,m),
\label{gen1Lint}
\ee
with $ a$, $b \geq 1$.
For general spin $j$ the formal expression
$$
   \langle T_j(x) T_j(y) \rangle_{c,0} =
   \hbar^2 \int {\dif^2 p  \over (2\pi)^2}  e^{i p\cdot(x-y)}
   \left[ (2j^2 -2j+1) I_{22} (p,m) + (2j^2 -2j) I_{31}(p,m) \right] \,
$$
is likewise converted to the well-defined
\bea
     &\displaystyle{
      \langle \norm{2}{T_j(x)} \norm{2}{T_j(y)} \rangle_{c,0} = }
\nonumber\\
     &\displaystyle{
  \hbar^2 \int {\dif^2 p  \over (2\pi)^2}  e^{i p\cdot(x-y)}  \left[
  (2j^2 -2j+1)\, (1 -t^2_p)\cdot I_{22} (p,m)
  + (2j^2 -2j)\, (1 -t^2_p) \cdot I_{31}(p,m) \right]\, .}&
\label{NOTTg}
\eea

The explicit computation of the BPHZ renormalized expression $I^R_{ab}(p)$
is  straightforward, and details are relegated to appendix\,\ref{apint}.
We simply quote the final result
in the limit $m \rightarrow 0$, namely \bref{Ren1Lint}
\be
  I^R_{ab}=-{i \over 2\pi}\, {p^{a+b-1} \over \bar p} B(a,b) \, ,
\ee
with $B(a,b)$ the Euler Beta function. The renormalized
correlation functions corresponding to (\ref{NOTTm}), (\ref{NOTTg})
are found to be
\bea
    \langle \norm{2}{T(x)} \norm{2}{T(y)} \rangle_{c,0} & = &
    {i  \hbar^2 D \over 24 \pi} \int {\dif^2 p \over (2\pi)^2}
    e^{i p\cdot (x-y)} {p^3 \over \bar p} \, ,
\nonumber\\
    \langle \norm{2}{T_j(x)} \norm{2}{T_j(y)} \rangle_{c,0} & = &
 {- i \hbar^2  \over 24 \pi}\, \left[2(6j^2 -6j +1) \right]
\int {\dif^2 p  \over (2\pi)^2}  e^{i p\cdot(x-y)}    {p^3 \over
\bar p} ,
\label{RNOTTg}
\eea
and
the renormalized expression for the lhs of (\ref{TT}) is
\be
\langle \norm{2}{T_t(x)} \norm{2}{T_t(y)} \rangle_{c,0} =
{i \hbar^2  \over 24 \pi} (D - 26)
\int {d^2 p  \over (2\pi)^2}  e^{i p(x-y)}    {p^3 \over \bar p}  \, .
\label{LHSTT}
\ee

After computing the rhs of \bref{TT} in  the
momentum representation in the limit $m\rightarrow0$
\be
    -i \hbar \alpha \langle \partial^3 c(x) b(y) \rangle_{c,0}
    = - \hbar^2 \alpha
    \int {d^2 p  \over (2\pi)^2} e^{i p\cdot(x-y)}{p^3 \over \bar p} \, ,
\label{RHSTT}
\ee
a comparison of \bref{LHSTT} and \bref{RHSTT} gives
$$
    \alpha = {-i \over 24 \pi} (D -26) \, .
%\label{int9}
$$
Combining this with the previous results about the functional form of
the anomaly, one finally obtains
\be
    \cA = {i (D-26)\over 24 \pi} \int d^2 x \,\,c\,\partial^3\,b^*\, ,
\label{anomW2}
\ee
a well-known result.

\subsection{The $W_2$ anomaly from Zimmerman identities}
\label{1law2mt}

\hspace{\parindent}%
In the previous subsection we showed the relation of the anomaly with
the Ward identity for correlation functions. In this subsection we show
its relation with the local operator equation \bref{anomeq2}.
It  may arise from two
different sources: normal ordering corrections originating
in the anisotropy of the variation of the interactions,
and  the  symmetry breaking produced
by the IR regulating mass term.
If we take both the interaction terms and
the transformations in \bref{W2action} to be minimally subtracted
normal products, only the second source contributes:
\bea
    &\displaystyle{
  - i \hbar \int \dif^2 x \,\, \norm{2}{\cA (x)}= }&
\nonumber\\
    &\displaystyle{ \lim_{m^2\rightarrow 0}\int \dif^2 x \,\, \left(
    \norm{2}{\frac{\dr S_m}{\delta\Phi^A(x) }
    \left\{\frac{\dl S}{\delta\Phi^*_A(x)}\right\} } -
    \norm{0}{\frac{\dr S_m}{\delta\Phi^A(x) }
    \frac{\dl S}{\delta\Phi^*_A(x)} } \right)\,.}&
\label{W2B1}
\eea
We have inserted the term with the minimally subtracted operator,
which is supposed to vanish in the limit $m \rightarrow 0$
anyway, to bring out the fact that classically this quantity is zero.
The  variation of the mass terms \bref{mama} and \bref{magh}
reads explicitly
\bea
    &\displaystyle{
    \norm{2}{ \frac{\dr S_m}{\delta\Phi^A(x)}
    \left\{\frac{\dl S}{\delta\Phi^*_A(x)}\right\} }=
    - m^2 \norm{2}{ \phi^i \left\{ c \partial \phi^i\right\}(x)+
     \left({1 \over \partial} c \right) \left\{ T \right\}(x) } }&
\nonumber\\
    &\displaystyle{
    + m^2 \norm{2}{ \left({1 \over \partial} b \right)
    \left\{(\partial c)c \right\}(x)
    + \left({1 \over \partial}c \right)
    \left\{ 2 b \partial c + (\partial b) c \right\}(x) } \, .}&
\label{nint10}
\eea

A non-zero result for \vgl{W2B1} arises from a difference in
the subtractions. We first spell out the influence of the anisotropy
on the various terms in \vgl{nint10}, and later turn to the influence of
the oversubtractions. For the terms involving matter fields,
when inserted in loop correlation functions%
\footnote{The only non-zero correlation functions that can be formed
with \bref{nint10} and elementary fields is
$\expect{\norm{}{\cA}(x) b(y)}_0$. See the previous subsection.}%
, the absence of interaction terms containing the product
$b\phi^i$ forces the factor $\phi^i$
to be {\it always} connected to an {\it internal} line, whereas the
factor $\left({1 \over \partial} c \right)$, instead, is always
connected to an {\it external} line. Both then produce matter
loops only. As a consequence of the definition
\bref{ani rule}, one sees that the first of these anisotropic
normal products behaves {\it effectively} as an isotropic, oversubtracted
one, whereas the second  reduces in fact to its isotropic, minimally
subtracted version (and thus produces no contribution to the anomaly~%
\bref{W2B1}). Similar conclusions can be drawn for the first
of the pure ghost terms of \vgl{nint10}.
Since the factor $\left({1 \over \partial} b\right)$  is necessarily
internal in a loop, the anisotropic normal ordering
effectively behaves once again as an isotropic oversubtracted
product.
For the second pure ghost term, the net effect of the anisotropy
is to first eliminate ghost loop contributions formed by considering the
factor $\left({1 \over \partial} c\right)$ as external, and to force
the remaining ones to be isotropic, although oversubtracted.

Having disentangled the consequences of the anisotropy,
the oversubtraction can now be taken into account.
This is done through the Zimmerman identity \bref{ZimId}
\be
    \norm{2}{\Theta(x)}- \norm{0}{\Theta(x)}=
    \sum_k \rho_k \norm{d_k}{\cM_k(x)},
\label{int12}
\ee
with $d[\cM_k]=d_k \leq 2$, and
with the composite operator $\Theta(x)$ given by
\bea
    &\displaystyle{
    \Theta(x)\equiv \Theta_{\rm m}(x) +\Theta_{\rm gh}(x)=}&
\nonumber\\
    &\displaystyle{
    - \phi^i \left\{c \partial \phi^i (x) \right\}
    +\left[ \left({1 \over \partial} b \right)
    \left\{(\partial c)c \right\}(x)
    + \left({1 \over \partial}c \right)
    \left\{ 2 b \partial c + (\partial b) c \right\}(x)\right].}&
\label{theta}
\eea
The coefficients are most easily determined by inserting
both sides of \bref{int12} into correlation
functions with arbitrary products of fields.
The insertion $\Theta(x)$ \bref{theta} has spin and dimension zero, which
restricts the normal product corrections on the rhs of
\bref{int12} to be exactly of the form
$\sum_k m^{-2}\tilde \rho_k\norm{2}{\Phi^*_A F^A_k}$,
with $\tilde\rho_k$ mass and coupling constant independent numerical
coefficients. The actual value of the
coefficients $\tilde \rho_k$ and  the field content of the objects
$F^A_k$ is again contained in
the loop contributions to the term linear in antifields, namely
\bea
   &\displaystyle{ \ihbar \int d^2 x \,\,\int d^2 y\,\,
   \expect{\norm{2}{\Phi^*_A(x) R^A(x)}
 \left\{ \norm{2}{\Theta(y)} - \norm{0}{\Theta(y)}  \right\}
               \prod_i \Phi^{A_i} (x_i)} _{c,0} =}&
\nonumber\\
  &\displaystyle{\sum_k \rho_k    \int d^2 x \,\,
  \expect{ \norm{2}{\Phi^*_A(x) F^A_k(x)}
   \prod_i \Phi^{A_i} (x_i)}_{c,0}}\, .&
\label{int14}
\eea
Considering possible double contractions
between $\Theta(y)$ and $R^A(x) = R^b(x)$, which always leave a
factor $c(y)$ free, the test product of fields is
again  $b(z)$ and the operator on the rhs can only be
$\Phi^*_A(x) F^A(x)= b^* \partial^3 c(x)$.
Differentiating w.r.t.\ the antifield $b^*$, \bref{int14} reduces to
\bea
    &\displaystyle{
   -\ihbar \int \dif^2 y\,\,
   \expect{\norm{2}{T_t(x)}
   \left\{ \norm{2}{\Theta(y)} - \norm{0}{\Theta(y)}  \right\}
    b(z)}_{c,0} =}&
\nonumber\\
    &\displaystyle{\rho
    \expect{\norm{2}{\partial^3 c(x)} b(z)}_{c,0}},&
\label{int15}
\eea
which is again an equation for the numerical
coefficient $\rho$ of the potential anomaly $\gh A= b^*\partial^3 c$.

Let us now evaluate the contribution $\rho_{\rm m}$ of the matter sector
--coming from double contractions between $\Theta_{\rm m}(y)$
\bref{theta} and $T(x)$-- to the coefficient $\rho$ in eq.\,\bref{int15}.
The BPHZ renormalized correlation function is
\bea
    \Lambda_a(x-y) & \equiv &
     \frac{i}{\hbar} \int \dif^2 z \,\, \expect{ \norm{2}{T(x)}\,
    \norm{a}{ \phi^i(z) c(z)(\partial\phi^i)(z)}\, b(y)}_{c,0}
\nonumber \\
    & = & \frac{\hbar^2 D}{2} \int {\dif^2 p  \over (2\pi)^2}
    e^{i p\cdot(x-y)} {p^2 \over p\bar p + m^2}
    (1 - t^a_p)\cdot I_{11}(p,m) \, .
\nonumber
\eea
The short hand notation \bref{shn} was used again.
The property
$t^{a+1}_p \, p= p\, t^{a}_p$ allows the commutation of
the explicit factor $p$ coming
from the $\partial c$ through the Taylor operator $t^{a+1}_p$,
that is
assigned by eq.\,\bref{j rule} to the 1PI part $p\, I_{11}(p,m)$ of the
above integral. For $a=0$ the new 1PI integral is seen to be
minimally subtracted --indeed, $I_{11}(p,m)$ is logarithmically divergent
(cf. appendix\,\ref{apint})-- while for $a=2$ it is oversubtracted.
We are interested in the difference, see \vgl{int15},
i.e.\,in the subtraction terms
themselves, of zero and second order in $p$:
\be
    \Lambda_2(x-y) - \Lambda_0(x-y) =
    \frac{\hbar^2 D}{2} \int {\dif^2 p  \over (2\pi)^2}
    e^{i p\cdot(x-y)} {p^2 \over p\bar p + m^2}
    (t^0_p - t^2_p)\cdot I_{11}(p,m).
\label{difference}
\ee
All terms are UV convergent, since they correspond to oversubtractions.
The integral can be evaluated without further ado
$$
    (t^0_p - t^2_p)\cdot I_{11}(p,m)
    = - m^2 p^2 \int {d^2 q \over (2\pi)^2}
    { q \bar q \over (q\bar q + m^2)^4} =
    {i \over 24 \pi} {2 \over m^2} \, p^2 \, ,
$$
providing the following result for the difference \bref{difference}
$$
    \Lambda_2(x-y) - \Lambda_0(x-y) =
    { i \hbar^2 D \over 24 \pi m^2}\int {\dif^2 p  \over (2\pi)^2}
    e^{i p\cdot(x-y)} {p^4 \over p\bar p + m^2} \, .
$$
After comparison with the rhs of \bref{int15} in  the
momentum representation,
\be
    \rho \langle \partial^3 c(x) b(y) \rangle_{c,0}
    = - i\rho\hbar
    \int {\dif^2 p  \over (2\pi)^2} e^{i p\cdot(x-y)}
    {p^4 \over p\bar p+m^2} ,
\label{mom rep}
\ee
the corresponding value for the contribution of the
matter sector to the coefficient $\rho$ in eq.\,\bref{int15} is read off:
$$
  \rho_{\rm m}= \frac{-\hbar D}{24 \pi m^2}.
$$

In the same way, the ghost sector contribution may now be evaluated
by considering the double contractions between $T_{\rm gh}(x)$ and
$\Theta_{\rm gh}(y)$ \bref{theta} in eq.\,\bref{int15},
taking the factor $\left({1 \over \partial} c\right)$  internal.
The relevant BPHZ renormalized correlation functions
$\tilde\Lambda_a(x-y)$ is
{ \setlength{\arraycolsep}{0pt}
\bea
\tilde \Lambda_a(x-y) &=&  \ihbar \int \dif^2 z\,
   \expect{ \norm{2}{ 2 b \partial c + (\partial b) c}(x)
   \norm{a}{ \left({1 \over \partial} b \right)
   \left\{ (\partial c)c \right\} +
   \left({1 \over \partial} c \right)
   \left\{ 2 b \partial c + (\partial b) c \right\} }(z) b(y) }_{c,0}
\nonumber\\
&=&- 2 \hbar^2 \int {\dif^2 p  \over (2\pi)^2}
    e^{i p\cdot(x-y)} {p \over p\bar p + m^2}
    (1-t^{a+1}_p) \cdot \tilde I(p,m),
\nonumber
\eea }
with the unrenormalized expression of the 1PI integral given by
\be
    \tilde I(p,m)= \int {\dif^2 q \over (2\pi)^2}\,
   { 2p^2 q + q^2 p -q^3 \over
   (q\bar q + m^2 )[(p-q)(\bar p - \bar q) + m^2 ]} \,.
\label{nint18}
\ee
The ghost sector contribution to the lhs of eq.\,(\ref{int15}) is
the difference
\be
    \tilde\Lambda_2(x-y) - \tilde\Lambda_0(x-y) =
    -2\hbar^2  \int {\dif^2 p  \over (2\pi)^2}
    e^{i p\cdot(x-y)} {p^2 \over p\bar p + m^2}
    (t^1_p - t^3_p)\cdot \tilde I(p,m).
\label{difference2}
\ee
Keeping in the Taylor series for the integrand in
\bref{nint18} only those numerators that survive the $q$-integration, namely
the ones of the form $(q\bar q)^n$, yields
$$
    (t^1_p-t^3_p)\cdot \tilde I(p,m)= \int {\dif^2 q \over (2\pi)^2}
    \left[ {2 q\bar q \over (q\bar q + m^2)^3} +
    { (q\bar q)^2 \over (q\bar q + m^2)^4} -
    { (q\bar q)^3 \over (q\bar q + m^2)^5} \right]=
    \frac{13 i}{24 \pi}\,\frac{p^3}{m^2}.
$$
The final result for \bref{difference2} is  then
$$
     \tilde \Lambda_2(x-y) -  \tilde \Lambda_0(x-y) =
    {-26 i \hbar^2\over 24 \pi m^2} \int {\dif^2 p  \over (2\ pi)^2}
    e^{i p\cdot(x-y)} {p^4 \over p\bar p + m^2} \, ,
$$
from which, by comparison with the momentum representation
\bref{mom rep} of the rhs of eq.\,\bref{int15}, the following value
for the ghost sector contribution to the coefficient $\rho$ in
eq.\,\bref{int15} is found
$$
  \rho_{\rm gh}= \frac{26 \hbar}{24 \pi m^2}.
$$

The results can be summarized by giving the Zimmerman
identity \bref{int12} explicitly as
$$
    \norm{2}{\Theta(x)}- \norm{0}{\Theta(x)}=
    \frac{ (26-D) \hbar}{24 \pi}
    {1 \over m^2} \,\norm{2}{b^*\partial^3 c(x)}.
$$
Inserted in eq.\,\bref{nint10}, it provides, after canceling
the $m^2$ factors, the following form for the
anomaly \bref{W2B1}
$$
   \cA = {i(D-26) \over 24 \pi} \int \dif^2 x \,\, c \partial^3 b^* ,
$$
in full agreement with the result computed in the previous subsection
\bref{anomW2}.

In conclusion, the local form of the anomaly equation enabled us to trace
back the source of the anomaly to the oversubtraction of the variation of
a mass term. The remainder of this evanescent
term in the massless limit was extracted through a Zimmerman identity.

\section{The anomaly of chiral $W_3$ gravity}
\label{W3}

\hspace{\parindent}%
In this section we first compute the anomaly of the chiral $W_3$ gravity
model \ct{W3} using the anomalous Ward identities (\ref{anomeq1}).
Afterwards we also show how the two-loop contribution arises from
(part of) the BRST breaking produced by the infrared regulating mass term,
again through oversubtractions and Zimmerman's identity.

$W$-gravity models have recently attracted some attention for
several reasons: as models with a non-linear symmetry algebra,
as possible higher spin extensions of gravity, etc.\
For our purposes, the basic interest in the $W_3$ model lies in the fact
that it is an example of a field
theory with a genuine new contribution to the anomaly at the two-loop
level. In other words, the rhs of (\ref{AZJeq}) contains at the two-loop
level not only nonlocal one-loop dressings of the one loop anomaly, but
extra {\it local} two-loop contributions to the anomaly.
The expression of the two-loop anomaly on the local level,
\vgl{anomeq2},
rather than through its insertion in the effective action, adds extra
clarity in that it is not necessary to disentangle it from the dressing
of the one-loop anomaly.
The expression we obtain for the anomaly in this model completely agrees
with the existing literature
\cite{mat89,hull91,ssn91,prs91,hull93,ST,roya,j95,jw95}.
Existing derivations of the two-loop contributions
always involved --implicitly or explicitly--
the use of OPE's of conformal field theory (for the one-loop
anomaly, derivations using traditional field theory methods
are found in \ct{ST,j95}). Our computation
makes no difference between one and higher loop anomalies.

\subsection{The $W_3$ model}

\hspace{\parindent}%
Chiral $W_3$ gravity is the minimal higher spin extension of chiral $W_2$
gravity, in which matter fields couple to gauge fields through
their spin 2 energy-momentum tensor $T$ and a spin 3 current~$W$
\be
  T=\frac12(\partial\phi^i)(\partial\phi^i),\quad\quad
  W=\frac13 d_{ijk} (\partial\phi^i) (\partial\phi^j) (\partial\phi^k)\, .
\label{currents}
\ee
The constants $d_{ijk}$ that determine the $W$ current are
totally symmetric  and satisfy
$$
  d_{i(jk}d_{l)mi}=k\delta_{(jl}\delta_{k)m},
$$
for some arbitrary, but fixed parameter $k$.

The extra gauge symmetry generated by the spin 3 current $W$ is
handled by supplementing
the  set of fields and antifields of chiral $W_2$ gravity with
an extra ghost-antighost pair $(u,v)$, and their associated antifields.
A tractable form of the extended action%
\footnote{See footnote~\ref{S en Sq}}
for chiral $W_3$ gravity in a
gauge fixed basis can   be taken from \ct{ST}%
\footnote{The free parameter $\alpha$ that appears in \ct{ST} is taken to
           be zero here.}
\bea
   S=\int\dif^2 x\hspace{-5mm}&&\left\{
   \left[-\frac12(\partial\phi^i)(\bar\partial\phi^i)
   +b(\bar\partial c)+ v(\bar\partial u)\right]\right.
\nonumber\\
   &&+\phi^*_i\left[c(\partial\phi^i)+
   u d_{ijk} (\partial\phi^j) (\partial\phi^k)
   -2 k b (\partial u) u (\partial\phi^i)\right]
\nonumber\\
   &&+b^*\left[-T +2 b(\partial c)+ (\partial b) c
   + 3 v(\partial u) + 2(\partial v) u \right]
\nonumber\\
   && +v^*\left[-W +2 k T b(\partial u) +2 k \partial(T b u)
   + 3 v(\partial c) + (\partial v) c \right]
\nonumber\\
   && \left.
   +c^*\left[ (\partial c) c
   +2 k T (\partial u) u\right]
   +u^*\left[2(\partial c) u- c(\partial u)\right]\right\}
\nonumber\\
     &=& S_0+\int\dif^2 x\,\, \Phi^*_A(x) R^A(x) \, .
\label{W3action}
\eea
The antifields $b^*$ and $v^*$ take over the role of (minus)
the gravitational gauge field and its spin 3 counterpart, respectively.
The expressions multiplying these antifields in
\bref{W3action} are then
(minus) the total energy momentum tensor $T_t$ and the total spin 3
current $W_t$. The extended action (\ref{W3action}) shares an
important property with (\ref{W2action}), namely they are both of the form
$S = S_0 + \int\norm{2}{\Phi^*_A R^A}$, with $S_0$ purely quadratic
in the fields: correlation functions for this model can therefore
again be  considered as perturbative series in the antifields.

The assignment of the relevant additive charges as in the
$W_2$ model (Table \ref{tbl:charges}), are now completed with Table
\ref{tbl:chargesII} for the extra fields.
%%%%%%%%%%%%%%%%%%%%%%%%%%%%%%%%%%%%%%%%%%%%%%%%%%%%%%%%%%%%%%%%%%%%%%%%%%%
\begin{table}[htf]
\begin{center}
\begin{tabular}{||c||c|c|c|c||}
\hline
 & dim $d$ & spin $j$ & gh.\,$\sharp$ & $d-j$  \\
\hline \hline
$u$& -2 & -2 & 1 & 0 \\
$v$  &  3 & 3& -1& 0 \\
\hline
$u^*$  & 4 & 2 & -2 & 2 \\
$v^*$ & -1 & -3 & 0 & 2 \\
\hline
    \end{tabular}
\caption{Additive charges for fields and antifields of the spin 3 sector.}
\label{tbl:chargesII}
\end{center}
\end{table}
%%%%%%%%%%%%%%%%%%%%%%%%%%%%%%%%%%%%%%%%%%%%%%%%%%%%%%%%%%%%%%%%%%%%%%%%%%%

We finish the specification of the model by choosing
the IR regulating mass term to be used in perturbative
computations. For the matter fields
the mass term (\ref{mama}) is used, such that the propagator is given by
(\ref{proma}), whereas both for the spin 2 $(b,c)$ ghosts and for the
new spin 3 $(u,v)$ ghosts a mass term of the form (\ref{magh}) is taken,
leading to a propagator of the form (\ref{progh}) for both ghost pairs.

\subsection{$W_3$ anomalous Ward identities}
\label{1lw3}

\subsubsection{Preliminary considerations}

\hspace{\parindent}%
We first analyze the anomalous Ward identities (\ref{anomeq1}),
as in the $W_2$ case, to find the general structure of the $W_3$ anomaly
and describe the strategy we follow for its computation.

As for the $W_2$ case, it follows from
Table \ref{tbl:charges} and Table \ref{tbl:chargesII}
and the specific antifield dependence of the action
that the BRST anomaly is linear in antifields, of the form
$\gh A=\Phi^*_A F^A$. The equation from which one can derive
information on the function $F^A$ is again of the form~%
(\ref{ord1bis}), in which the quantities $R^A$ should now be read off from
the extended gauge-fixed action~\bref{W3action}.
Since loop diagrams come from at least double contractions
between $R^A(x)$ and $R^B(y)$ on the lhs of \bref{ord1bis},
the fact that  $R^A$ contains terms  with up to four fields
indicates that relevant diagrams for the computation of the anomaly may
contain up to three loops.
However, a closer look at the
$\gh O(\Phi^4)$ terms in the $R^A$'s, given by
\be
    -2 k b (\partial u) u (\partial \phi^i),\quad\quad
     2k \left[ T b \partial u + \partial (Tbu) \right],\quad\quad
     2k \left[ T (\partial u) u \right],
\label{int30}
\ee
and at the form of the propagators,  shows that
no {\em proper} three-loop diagram can arise formed by  quartic
contractions between the expressions in (\ref{int30}).
The only two loop diagram arises
from a triple contraction between two matter spin 3 currents $W$
\bref{currents},
so that the potential two-loop anomaly  depends only on the antifield
$v^*$. Finally, double contractions between the different terms in
the $R^A$'s give rise to various one-loop
contributions to the anomaly. Those are described in detail below.
In summary, only one and two loop anomalies are possible in this model.

The antifield coefficients $F^A(\Phi)$ determining these
one and two loop contributions to the anomaly are determined as in
the $W_2$ case. First one selects those pairs $R^A(x)$, $R^B(y)$
that admit double and triple contractions,
and in a second step determines the ``test product'' of fields
to be contracted with the remaining external legs.
The precise form of the test product and a subsequent
dimensional analysis completely determines the fields
and the {\it number} of $\partial$-derivatives  present
in the coefficient $F^A$. The precise derivative structure
and the numerical coefficients are obtained
from an explicit (renormalized) computation of the lhs of (\ref{ord1bis}).
In doing this last step, the minimal normal ordering of the interactions
prescribes minimal subtractions for all the divergent one loop integrals,
which should therefore be substituted by their minimally renomalized
counterpart \bref{Ren1Lint}. BPHZ renormalization of the relevant two-loop
integral, requiring the use of Zimmerman's forest formula, will
instead be treated {\it in situ}.

\subsubsection{One loop anomaly: a sample computation}
\label{1lsample}

\hspace{\parindent}%
In this subsection, the procedure is illustrated by
computing the ``gravitational'' part to the complete one loop BRST
anomaly, viz. the contribution proportional to the antifield $b^*$.
This suffices to get a flavour of the method.
Further details on the computation of the full one loop anomaly can be
found in appendix \ref{w31lcom}.

Three different contributions to this gravitational part can be discerned:
i) double contractions of $R^b=-T_t$ with itself;
ii) contributions containing one factor $d_{ijk}$; and
iii) contributions proportional to $k$, containing two such factors.
We discuss them in detail in this order.

\medskip

{\bf i) }\hspace{1em}
This is analogous to the computation in section~\ref{W2}.
One uses
$R^b = - (T + T_2 + T_3)$, with $T_j$ defined by (\ref{Teejay}) with
suitable substitution of $(b,c)$ by $(u,v)$ in $T_3$. Going  through
exactly the same steps as in sect.\,\ref{w2awi}, but taking two
contributions of the type (\ref{RNOTTg}) into account, namely for $j=2$
and for $j=3$, this first contribution to the
one-loop anomaly of chiral $W_3$ gravity is
\be
    \cA_1^{(i)}  = {i (D-100)\over 24 \pi}
    \int \dif^2 x \,\, c\, \partial^3 b^*
\label{an11} \, .
\ee

\begin{figure}
\epsfbox{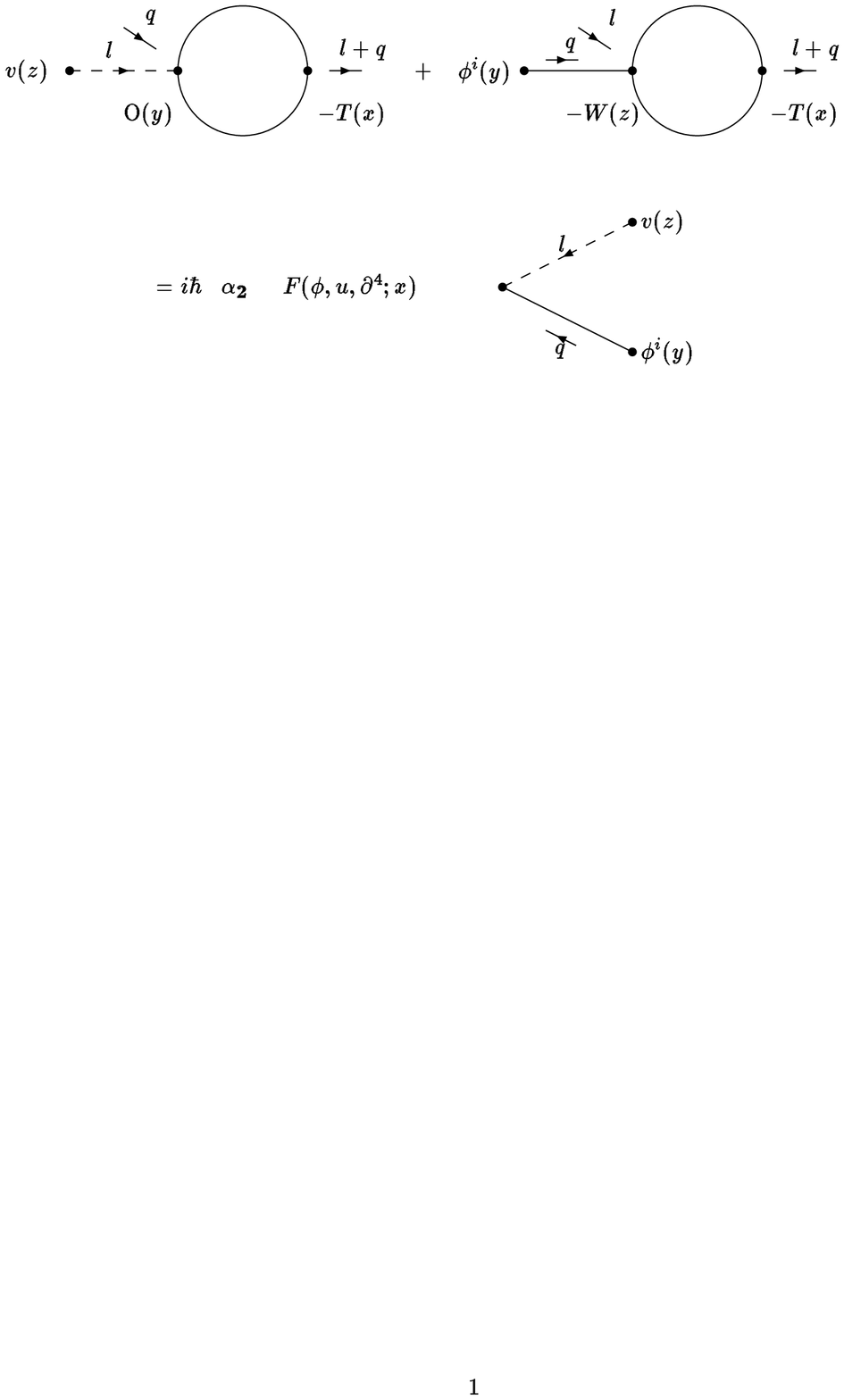}
\caption{One-loop anomalous diagrams for $W_3$. The insertion in the first
diagram is
$O(y)=[ud_{ijk}(\partial \phi^j)(\partial \phi^i)](y)$
}\label{W3-1loop}
\end{figure}

{\bf ii) }\hspace{1em}
Contributions to the
$b^*$-anomaly proportional to the tensor $d_{ijk}$
are generated by loops of matter fields coming from
double contractions of $-T$ in $R^b$ with the terms
$$
  \left\{ \begin{array}{ccc}
   u d_{ijk} (\partial\phi^j) (\partial\phi^k) & \mbox{ in }  & R^{\phi} \\
  - \frac13 d_{ijk} (\partial\phi^i) (\partial\phi^j) (\partial\phi^k)
   & \mbox{ in } & R^v \, .
   \end{array} \right.
$$
Obviously, a suitable test product for both contributions
is $\phi^i(y) v(z)$, so that this part of the anomaly  has to
contain one factor $\phi^i$ and one factor $u$, while dimensional arguments
require four derivatives $\partial$. For this contribution, the
Ward identity (\ref{ord1bis}) takes the form
\bea
  &\displaystyle{
  \ihbar \expect{ \norm{2}{-T(x)} \norm{0}{u d_{ijk} (\partial\phi^j)
  (\partial\phi^k)(y)} v(z)}_{c,0}}&
\nonumber\\
  &\displaystyle{
  +\ihbar \expect{ \norm{2}{-T(x)}
  \norm{3}{ \frac13 d_{jkl}(\partial\phi^j)
  (\partial\phi^k)(\partial\phi^l)(z)} \phi^i(y)}_{c,0} }&
\nonumber\\
  &\displaystyle{ = - \alpha_2
\expect{ \norm{2}{F(\phi,u,\partial^4;x)} \phi^i(y) v(z)}_{c,0},}&
\label{int32}
\eea \newline
which can be represented by figure~\ref{W3-1loop}.
These  one--loop contributions are
essentially the same as for the $W_2$. They are minimally
subtracted, i.e. up to quadratic terms in the external
momentum (see \vgl{j rule}).
The resulting finite integrals are computed in
\vgl{Ren1Lint}, and the lhs of (\ref{int32}) becomes
\bea
  &&{\hbar^2 \over 12 \pi} d_{ill} \int {\dif^2 l \over (2\pi)^2}
   \int {\dif^2 q \over (2\pi)^2} e^{- i l \cdot z} e^{- i q \cdot y}
   e^{i(l+q) \cdot x}
   \left[ {(l+q)^3 \over (\bar l+ \bar q)} {1 \over\bar l}+
   {(l+q)^3 \over (\bar l+ \bar q)}{1\over \bar q}\right]=
\nonumber\\
  &&{\hbar^2 \over 12 \pi} d_{ill} \int {\dif^2 l \over (2\pi)^2}
   \int {\dif^2 q \over (2\pi)^2} e^{- i l \cdot z} e^{- i q \cdot y}
   e^{i(l+q) \cdot x} {(l+q)^3 \over \bar l \bar q}   \, .
\nonumber
\eea
This momentum dependence leads to the conclusion that in \vgl{int32}
$F = \partial^3 ( u (\partial \phi^i) d_{ill} )$, and determines the
coefficient $\alpha_2$. This yields the
second contribution to the anomaly:
\be
   \cA^{(ii)}_1 = {i \over 24 \pi}  \int \dif^2 x \,\,
   [2 u (\partial \phi^k)  d_{ijk}] \partial^3 [b^* \delta^{ij}] \, .
\label{an12}
\ee

\medskip

{\bf iii) }\hspace{1em}
Contributions to the
$b^*$-anomaly proportional to $k$
are generated by matter loops coming from two contractions of $-T$ in
$R^b$ with the terms
$$
    \left\{
    \begin{array}{ccc}
    2 k T b(\partial u) +2 k \partial(T b u)
    & \mbox{ in }  & R^v \\
    2 k T (\partial u) u & \mbox{ in } & R^c \, .
   \end{array}   \right.
$$
A suitable test product for both contributions is now
$v(y) v(z) c(t)$, yielding a possible anomaly containing
two $u$ fields, one $b$ field and four derivatives $\partial$ in order to
match the dimensions. (\ref{ord1bis}) reads now
\be
\begin{array}{c}
\{   \ihbar \expect{ \norm{2}{T(x)}
\norm{3}{2 k T b(\partial u)(y) +2 k \partial(T b u)(y)} v(z) c(t)}_{c,0}
- (y \leftrightarrow z) \} \\
   \\
   +\ihbar \expect{ \norm{2}{T(x)} \norm{-1}{2 k T (\partial u) u(t)}
   v(y) v(z) }_{c,0} \\
   \\
 = - \alpha_3 \expect{ \norm{2}{ F(b,u,u,\partial^4;x)} v(y) v(z)
     c(t)}_{c,0}. \end{array}
\label{int35}
\ee
The subtraction degrees computed from the $N_a$,
once more the minimal ones, namely $2$, yield a renormalized
expression for the lhs of (\ref{int35}):
$$
  {\hbar^3 i k D \over 12 \pi} \int {\dif^2 l \over (2\pi)^2}
   \int {\dif^2 q \over (2\pi)^2}  \int {\dif^2 r \over (2\pi)^2}
    e^{ i (q + l + r) \cdot x} e^{- i q \cdot y} e^{- i r \cdot t}
    e^{- i l \cdot z}
    { (q + l + r)^3 (q-l)\over \bar q \bar l \bar r} \, .
$$
The momentum dependence determines $F = \partial^3 ( b u \partial u )$.
After computing
the numerical coefficient $\alpha_3$, the third contribution to the
anomaly is
\be
   \cA^{(iii)}_1 =
   { i \over 24 \pi} \int \dif^2 x \,\,
   [-2kb (\partial u) u \delta^{ij}] \partial^3[b^* \delta_{ij} ]\,.
\label{an13}
\ee

\medskip

\underline{In summary},
 collecting (\ref{an11}), (\ref{an12}) and
(\ref{an13}), the gravitational part of the one-loop $W_3$ anomaly is
\be
   \cA^{b^*}_1 = \cA^{b^*}_{1, \rm m} + \cA^{b^*}_{1, \rm gh} =
   { i \over 24 \pi} \int \dif^2 x \,\,  c^{ij}
     \partial^3 ( b^* \delta_{ij} ) - { 100 i \over 24 \pi}
      \int \dif^2 x \,\, c \,\partial^3 b^*,
\label{an1b}
\ee
where we have defined
\be
    c^{ij} = [ c - 2 k b (\partial u) u ] \delta^{ij} + 2 u
              d^{ij}_{\,\,\,k} (\partial \phi^k)  \,
\label{caaijee}
\ee
for further use.
The subscripts m and gh will stand from now on for purely
matter induced contributions and contributions induced by loops involving
(one or two) ghost lines, respectively.

This procedure
is applied time and again in appendix \ref{w31lcom}
in order to find the complete one-loop BRST anomaly.
The final result is
\be
  \gh A_1=\gh A_{1, \rm m}+ \cA^{b^*}_{1, \rm gh}
  +\cA^{\phi^*}_{1, \rm gh}+ \cA^{v^*}_{1, \rm gh}\, ,
\label{c1lw3}
\ee
with $\cA^{b^*}_{1, \rm gh}$, $\cA^{\phi^*}_{1, \rm gh}$ and
$\cA^{v^*}_{1, \rm gh}$ given by \bref{an1b}, \bref{an15} and \bref{an19}
respectively, and where $\gh A_{1, \rm m}$ lumps together all the matter
induced contributions in the compact expression
$$
   \gh A_{1, \rm m}=\frac{i}{24\pi} \int\dif^2 x\,
   c^{ij}\,\partial^3\, h^*_{ij},
$$
with the ``effective'' antifield $h^*_{ij}$ given by
$$
   h^*_{ij}=\left\{ \delta_{ij}
   \left[ b^* + 2kb(u(\partial v^*)-v^*(\partial u))
   +2k c^*u(\partial u)\right]
   -2 d_{ij}^{\,\,\,\,k}
   \phi^*_k u +2 v^* d_{ijk} (\partial\phi^k)\right\}.
$$
Our result for the one-loop
anomaly in the $W_3$ model is in perfect agreement with
previous computations in the literature
\cite{hull91,ssn91,prs91,hull93,ST,roya,j95}.

\subsubsection{Two-loop anomalous Ward identity}
\label{tlawi}

\hspace{\parindent}%
After this brief discussion of the one loop anomaly, we now turn
to the two loop anomaly. We consider it to be a major challenge
for the present setup, since, as far as we know, no computation has been
given that uses only the methods of renormalised perturbation theory.

As already pointed out, the unique two-loop diagram
constructed from three contractions between two terms from the available
$R^A$ is the diagram that corresponds to a
triple contraction between two copies of the spin-3 current $W$. This
immediately leads to a two loop anomaly proportional to $v^*$, since
in the general Ward identity (\ref{ord1bis}) the $R$'s are simply reduced
to $W$. A suitable test product is hence $v(y)$, which on the rhs yields a
function $F$ of the form $\alpha\partial^5 u$,
the only expression with non-vanishing correlation function with the
test product $v$ and with the correct dimension $2$. The Ward identity
(\ref{ord1bis})
takes the form
\be
   \ihbar \expect{\norm{3}{W(x)}\norm{3}{W(y)}} = \alpha
   \expect{\norm{3}{\partial^5 u(x)} v(y)} \, .
\label{int50}
\ee
This is the analogue of \bref{TT} in
chiral $W_2$ gravity and corresponds to the  triple contraction
in the standard conformal field theory OPE of $W(x)W(y)$.

The computation of the lhs of \bref{int50}, taking into account the
appropriate number of subtractions,
is a straightforward application of the forest formula. It is given
explicitly in \vgl{int61}. For readers unfamiliar with the rationale behind
this forest, we build up the expression by considering the regions
in momentum space which cause a divergence in the unsubtracted diagram.

\begin{figure}
\epsfbox{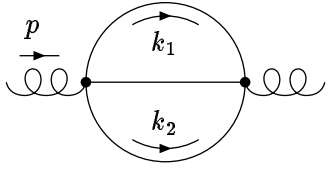}
\caption{The anomalous two-loop diagram.\label{2loop}}
\end{figure}

\begin{figure}
\epsfxsize=350pt
\centerline{\epsfbox{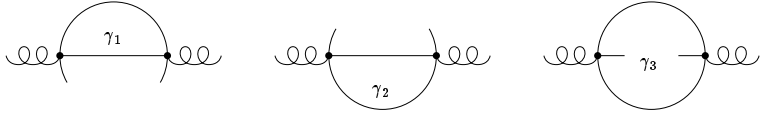}}
\caption{The 3 one-loop divergent subgraphs.\label{2loopcuts}}
\end{figure}

The unregularized version of the two-loop integral
$I\!\!I^R(p,m)$ is given by
\be
   I\!\!I(p,m) =  \int {\dif^2 k_1 \over (2\pi)^2} \int
   {\dif^2 k_2 \over (2\pi)^2}\,\,
   \cI\!\!\cI (k_1,k_2; p, m),
\label{int60c}
\ee
with the integrand %$\cI\!\!\cI (k_1,k_2; p, m)$ defined as
\be
   \cI\!\!\cI (k_1,k_2; p, m)=
   {k_1^2 \over k_1 \bar k_1 + m^2}\,
   {k_2^2 \over k_2 \bar k_2 + m^2}  {(p-k_1-k_2)^2\,
   \over \left[(p-k_1-k_2)(\bar p - \bar k_1 - \bar k_2) +m^2\right]}
   \, .
\label{int60}
\ee
\newline
The momentum labeling is as in figure~\ref{2loop}.
In terms of the corresponding properly subtracted expression,
the lhs of \vgl{int50} is
\be
    -\frac23 d^2
    \hbar^2 \int {\dif^2 p \over (2\pi)^2}
    e^{ i p \cdot (x-y)} I\!\!I^R(p,m) \, , \quad\quad
    \mbox{\rm with}\quad\quad d^2\equiv d^{ijk} d_{ijk}.
\label{int52}
\ee
Thus we have to determine that expression.
We start by identifying the possible superficial divergences.
There are four regions in the integration domain where the integral
potentially diverges:
\be
\left\{
  \begin{array}{ll}
     k_1 \rightarrow \infty,\,\, k_2 \mbox{ finite} & \mbox{quadratic
     divergence} \\
     k_2 \rightarrow \infty,\,\, k_1 \mbox{ finite} & \mbox{quadratic
     divergence} \\
     k_1 \rightarrow \infty,\,\, k_2 \rightarrow \infty,\,\,
     k_1+k_2 \mbox{ finite} & \mbox{quadratic divergence} \\
     k_1 \rightarrow \infty,\,\, k_2 \rightarrow \infty &\mbox{quartic
     divergence.}
  \end{array}
\right.
\label{int domains}
\ee
The quadratically divergent regions are associated
with the one-loop subdiagrams $\gamma_i$, $i=1,2,3$
--pictorially represented in figure~\ref{2loopcuts}--
obtained by cutting one of the three internal lines between $W(x)$ and
$W(y)$ in the original two-loop diagram (we take $\gamma_1$ to correspond
to cutting the $k_2$ line, $\gamma_2$ the $k_1$ line, and
$\gamma_3$ the $(p - k_1 -k_2)$ line).
The number of subtractions is, as always, determined from (\ref{j rule}),
and amounts to the minimal subtraction
of the quadratic subdivergences identified above.
The cut line  corresponds to a field $\partial\phi$,
and if we include the resulting momentum dependence of the numerator
(but not the propagator) in the expresssion for the subdiagram,
the subtraction degree is $4$.
The quartic divergence of the complete diagram $G$ also has
subtraction degree $4$.
Zimmerman's forest formula \cite{z} then consists of
two steps.
In the first step, the superficial quadratic divergences in
the original integrand associated
with the subdiagrams $\gamma_i$ are subtracted:
$$
   \cI\!\!\cI - \sum_{i = 1}^3 t^4_{p_{\gamma_i}}\cI\!\!\cI(\gamma_i),
$$
where $p_{\gamma_i}$ stands for the external momenta
of the $\gamma_i$ subdiagram and $\cI\!\!\cI(\gamma_i)$ for the suitable
restriction of the original integrand \bref{int60} to the subdiagram
$\gamma_i$.
This  leads to an expression that is
still quartically divergent. A further
quartic subtraction cures this:
$$
   (1-t^4_{p}) \left[\cI\!\!\cI
   - \sum_{i = 1}^3 t^4_{p_{\gamma_i}}\cI\!\!\cI(\gamma_i)\right].
$$

The net result of this process is Zimmerman's forest formula
adapted to our integrand \bref{int60}
\be
   \cI\!\!\cI - t^4_{p} \cI\!\!\cI - \sum_{i = 1}^3
   t^4_{p_{\gamma_i}}\cI\!\!\cI + \sum_{i = 1}^3 t^4_{p}
   t^4_{p_{\gamma_i}}\cI\!\!\cI
   = \sum_{\gh U\in \gh F}\,\prod_{\gamma\in \gh U}\,
   \left(-t^{d(\gamma)}_{p_\gamma}\right)\, \cI\!\!\cI,
\label{int61}
\ee
where $\gh F$ stands for the set of all possible forests $\gh U$ of the
two--loop diagram, namely
$$
  \gh F=\left\{ \emptyset,\{ G \}, \{ \gamma_1 \},
   \{ \gamma_2 \}, \{ \gamma_3 \}  ,\{ \gamma_1, G \},
   \{ \gamma_2,G \} ,\{ \gamma_3,G \}\right\}.
$$
Expression \bref{int61} is the subtracted integrand which upon
substitution in \bref{int60c} yields the proper expression of
our two-loop integral.

A closer inspection of \bref{int61} reveals that in its
last two sums the terms corresponding to $i=1,2$ cancel each
other. We exemplify this for $\gamma_1$, i.e.
$(1-t^4_{p})t^4_{p_{\gamma_1}}\cI\!\!\cI$. The expression of
the integrand \bref{int60} restricted to the subdiagram $\gamma_1$ reads
$$
   \cI\!\!\cI(\gamma_1) =
   {r_2^2 \over k_2 \bar k_2 + m^2}\,
   {k_1^2 \over k_1 \bar k_1 + m^2}\,
   {(p+r_1-k_1)^2
\over \left[(p+r_1-k_1)(\bar p + \bar r_1 - \bar k_1) +m^2\right]}\, ,
%\label{int62}
$$
where in the set of external momenta, $p_{\gamma_1}=\{p,r_1\}$,
$r_1$ must be identified with $-k_2$ {\it after} having done the
 Taylor expansion. In this way, the contribution generated by the
forest $\{ \gamma_1 \}$ is
\be
    t^4_{p_{\gamma_1}}\cI\!\!\cI(\gamma_1)
    \vert_{r_1 = -k_2} =
   {k_2^2 \over k_2 \bar k_2 + m^2}\, {k_1^2 \over k_1 \bar k_1 + m^2}\,
   \gh O(p^2),
\label{int63}
\ee
where $\gh O(p^2)$ stands for a polynomial of degree two in the momentum
$p$ whose coefficients are rational functions of $k_1$ and $k_2$. It is
then evident that the action of $(1-t^4_{p})$ on \bref{int63} produces a
vanishing result, since on quadratic polynomials in $p$, $t^4_{p}$ acts
like the identity. The same mechanism can be seen to hold as well for the
subdiagram $\gamma_2$, obtained from $\gamma_1$ upon
interchange of $k_1$ and $k_2$.

For $\gamma_3$ the situation is slightly different.
Indeed, the Taylor series expansion in the external momenta
$p_{\gamma_3}=\{p,r_3\}$ of the integrand \bref{int60} adapted to
$\gamma_3$, i.e.
$$
   \cI\!\!\cI(\gamma_3) =
   {k_2^2 \over k_2 \bar k_2 + m^2}\,
   {r_3^2 \over \left[(p+r_3-k_2)(\bar p + \bar r_3 - \bar k_2)
   +m^2\right]}\, {(p+r_3-k_2)^2
\over \left[(p-k_1-k_2)(\bar p - \bar k_1- \bar k_2) +m^2\right]}\, ,
$$
yields a nonpolynomial dependence on the external momentum $p$, namely
\bea
    &\displaystyle{t^4_{p_{\gamma_3}}\cI\!\!\cI(\gamma_3)
    \vert_{r_3 =(-p + k_1+k_2)} =}&
\nonumber\\
    &\displaystyle{
   {k_2^2 \over k_2 \bar k_2 + m^2} \,
   {(p-k_1-k_2)^2 \over
   \left[(p-k_1-k_2)(\bar p - \bar k_1 - \bar k_2) +m^2\right]}\,
    \restric{t^2_{p_{\gamma_3}}
   \left[{(p_{\gamma_3}-k_2)^2 \over
    \left[(p_{\gamma_3}- k_2)(\bar p_{\gamma_3}- \bar k_2)+m^2 \right]}
    \right]} {p_{\gamma_3} =(k_1+k_2)}.}&
\nonumber
\eea
This produces  a nonvanishing result when the operator
$(1-t^4_p)$ acts on it.
Using the more convenient integration
variables $l=k_1+k_2$ and $k= k_2$,
the renormalized expression of the two-loop integral
\bref{int60c} is thus found to be
\bea
   I\!\!I^R(p,m) &=&  \int {\dif^2 l \over (2\pi)^2}
   \left( 1 - t^4_p \right) {(p-l)^2
   \over \left[(p-l)(\bar p - \bar l)+m^2 \right]}\times
\nonumber\\
   &&\int {\dif^2 k \over (2\pi)^2}\,
   \left( 1 - t^2_l \right)\left[ {k^2 \over k \bar k + m^2}\,
   {(l-k)^2 \over \left[ (l-k) (\bar l-\bar k)+ m^2\right]}\right]\,.
\label{int66}
\eea

It is now a straightforward task to evaluate \bref{int66}. Indeed, since
the $k$--``one-loop'' integral is precisely of the minimally
subtracted form (\ref{1lapp},\ref{2lapp}) considered in appendix
\ref{apint}, with $a=b=2$, use of (\ref{MasRen1Lint},\ref{Ren1Lint})
brings it to the form
\be
  I\!\!I^R(p,m) = {-i \over 12 \pi} \int {\dif ^2 l \over (2\pi)^2}
  \left( 1 - t^4_p \right) \left[ F\left(\frac{m^2}{l\bar l}\right) \,
  {l^4 \over l\bar l + m^2}\,
  {(p-l)^2 \over \left[(p-l)(\bar p - \bar l) \right] + m^2}\right] \, .
\label{int67}
\ee
Apart from the extra factor
$F\left(\frac{m^2}{l\bar l}\right)$, with $F(0)=1$, this is of the
same generic minimally subtracted form (\ref{1lapp},\ref{2lapp}),
now with $a=4$, $b=2$. Therefore, in the limit of interest
$m\rightarrow 0$, the one-loop result (\ref{Ren1Lint}) can again be used,
yielding for \bref{int67}
$$
  I\!\!I^R(p,0) = - {1 \over 480 \pi^2} {p^5 \over \bar p} \, ,
$$
and for the lhs \bref{int52} of the Ward identity \bref{int50} the final
result
\be
    \frac {d^2 \hbar^2}{720\pi^2} \int {\dif^2 p \over (2\pi)^2}
    e^{ i p \cdot (x-y)} {p^5 \over \bar p} \, .
\label{int53}
\ee

Comparing the expression \bref{int53} with the rhs of (\ref{int50})
$$
   i \alpha \hbar \int {\dif^2 p \over (2\pi)^2} e^{ i p \cdot (x-y)}
   {p^5 \over \bar p} \, ,
$$
the value of the coefficient $\alpha $ is read off, and
 the complete form of
the $W_3$ two-loop anomaly is
\be
   \cA_2 = {i \hbar d^2 \over 720 \pi^2} \int \dif ^2x \,\,
    u \,\partial^5 v^* \, ,
\label{an2}
\ee
in agreement with \cite{ssn91,hull93,jw95}. This concludes our analysis
of the one- and two-loop anomalies in the $W_3$ gravity model by means of
the anomalous Ward identities.

\subsection{The $W_3$ two-loop anomaly from the mass term}
\label{2law3mt}

\hspace{\parindent}%
We show in this section
 how the two-loop contribution \bref{an2} to the
BRST anomaly for $W_3$ arises as a consequence of the explicit BRST
symmetry breaking by the IR regulating mass term, i.e.\ from the
eq.\,\bref{W2B1}. The main tool is again the reduction of the
oversubtracted BRST variation of the mass term through the use of
Zimmerman identities.

The $W_3$ IR regulating mass term is formed by adding
\bref{mama}, \bref{magh} and a copy of the latter for the $uv$ ghost
system. Its BRST variation is
\bea
    &\displaystyle{
    \norm{2}{\frac{\dr S_m}{\delta\Phi^A(x)}
    \left\{\frac{\dl S}{\delta\Phi^*_A(x )}\right\} }= }&
\nonumber\\
    &\displaystyle{
   - m^2 \norm{2}{ \phi^i(x) \left\{c(\partial\phi^i)+
     u d_{ijk} (\partial\phi^j) (\partial\phi^k)
     -2 k b (\partial u) u (\partial\phi^i) \right\} }}&
\nonumber\\
    &\displaystyle{
    + m^2 \norm{2}{ \left({1\over\partial}b\right) \left\{ R^c \right\}
    + \left( {1 \over \partial} c \right) \left\{ R^b \right\}
    + \left( {1 \over \partial} v \right) \left\{ R^u \right\}
    + \left( {1 \over \partial} u \right)
      \left\{ -W -W_{\rm gh}\right\}}\, .}&
\label{W3B1}
\eea
When inserted in loop correlation
functions, it gives rise to a multitude of one-loop diagrams
upon double contractions with the interaction terms $\Phi^*_A R^A$ in
\bref{W3action}. These diagrams are of course the analogues of the one-loop
correlation functions studied in sect.\,\ref{1lsample} and
in appendix \ref{w31lcom}.
We will not dwell on a rederivation of the one loop contribution
to the $W_3$ anomaly, but concentrate instead on the  more interesting
two-loop contribution in this approach.
Just as before,
two-loop contributions can only be generated by triple matter
contractions.
These occur now
between the interaction term $- v^* \norm{3}{W}$ and the
oversubtracted $\phi^3$ type terms in \bref{W3B1}, namely
\be
  - m^2 \norm{2}{ \phi^i \left\{u d_{ijk} (\partial\phi^j)
  (\partial\phi^k) \right\} },
  \quad\quad\mbox{and}\quad\quad
  - m^2 \norm{2}{\left({1 \over \partial} u \right) \left\{W\right\}}\, .
\label{diverse terms}
\ee
Both contractions leave a factor $u$ free and the
two-loop anomaly is therefore necessarily of the form
$\gh A_2= m^{-2}\tilde\rho v^*\partial^5 u$, in accordance with the
analysis presented in sect.\,\ref{tlawi}.
Anisotropy forces the second term in
\bref{diverse terms} to behave effectively
as a minimally subtracted, isotropic normal product,
since the factor $\partial^{-1} u$ is always in an external line.
Upon substitution in eq.\,\bref{W2B1}, no
two-loop contribution is obtained from this term.
The two-loop anomaly can then be extracted from
the  the Zimmerman identity
\be
  \norm{2}{\phi^i\left\{u d_{ijk}(\partial\phi^j)
  (\partial\phi^k)\right\}(x)}
  - \norm{0}{\phi^i\,u d_{ijk}(\partial\phi^j) (\partial\phi^k)(x)}=
   \rho\norm{2}{v^*\partial^5 u}+ \ldots.
\label{w3zid}
\ee
More specifically, the relevant coefficient will arise from a
two-loop calculation.
We define%
\footnote{Note that for minimally subtracted normal products,
          anisotropic and isotropic versions coincide.}
\be
   \Sigma_a  (x-y) = - \ihbar \int\dif^2 z \expect{ \norm{3}{W(x)}
    \norm{a}{ \phi^i \left\{u d_{ijk} (\partial\phi^j)
    (\partial\phi^k) \right\}(z) } v(y) }_{c,0} \, .
\label{W3B3}
\ee
The coefficient $\rho$ in \bref{w3zid}
is then determined by
\be
   \Sigma_2(x-y)- \Sigma_0(x-y)=
    \rho \expect{ \norm{3}{\partial^5 u(x)}v(y)}_{c,0}.
\label{sigma dif}
\ee
The unsubtracted 1PI two-loop integral corresponding to
\bref{W3B3}, is
\be
    2 i \hbar^3 d^2  \int {\dif^2 p \over (2\pi)^2}
    e^{ i p \cdot (x-y)} {p \over p\bar p + m^2} \widetilde{ I\!\!I}(p,m),
\label{W3B4}
\ee
where the integrand is
\bea
    &\displaystyle{
   \widetilde{I\!\!I}(p,m) =\int {\dif^2 k_1 \over (2\pi)^2}
   \int {\dif^2 k_2 \over (2\pi)^2}\,\,
   \widetilde{ \cI\!\!\cI }(k_1,k_2; p, m)}&
\nonumber\\
    &\displaystyle{
   = \int {\dif^2 k_1 \over (2\pi)^2}
    \int {\dif^2 k_2 \over (2\pi)^2}\,
    {k_1^2 \over k_1 \bar k_1 + m^2}\,
    {k_2^2 \over k_2 \bar k_2 + m^2}  {(p-k_1-k_2)
    \over \left[(p-k_1-k_2)(\bar p - \bar k_1 - \bar k_2) +m^2\right]}
    \, , }&
\label{W3B5}
\eea
which differs from the previous two-loop integral \bref{int60} in
one factor $(p-k_1-k_2)$. This changes the analysis of the
divergent integration domains slightly in comparison with
\bref{int domains}. Using the same momentum labeling, one has
\be
\left\{
  \begin{array}{ll}
     k_1 \rightarrow \infty,\,\,k_2 \mbox{ finite} & \mbox{linear
     divergence} \\
     k_2 \rightarrow \infty,\,\, k_1 \mbox{ finite} & \mbox{linear
     divergence} \\
     k_1 \rightarrow \infty,\,\, k_2 \rightarrow \infty,\,\,
     k_1+k_2 \mbox{ finite} & \mbox{quadratic divergence} \\
     k_1 \rightarrow \infty,\,\, k_2 \rightarrow \infty
     &\mbox{cubic divergence.}
  \end{array}
\right.
\label{W3B6}
\ee
The same labeling for the three different subdiagrams as in the previous
subsection is used. In particular, $\gamma_3$ corresponds to the quadratic
divergence in \bref{W3B6}. The subtraction  for the full two-loop
diagram is always isotropic,
and the degree is $2 + (3 - 2) + (a - 2) + (3 -1) = 3 + a$.
For the subdiagrams $\gamma_1$ and $\gamma_2$, the $\phi^i$ is
always in the loop itself, and the subtraction degrees are the same as for
the two-loop diagram. For the third
one-loop subdiagram $\phi^i$ is an external line, and
the prescribed subtraction is always the minimal subtraction to order $3$,
for both values of $a$. A
straigtforward application of Zimmerman's forest formula on the
integral (\ref{W3B4},\ref{W3B5}) yields for the integrand determining the
difference $\Sigma_2-\Sigma_0$ \bref{sigma dif} the symbolic expression
\be
(t^3_p - t^5_p) (1 - t^3_{p_{\gamma_3}})\cdot \widetilde{ \cI\!\!\cI }
+ \sum_{i=1,2} t^5_p (1 - t^5_{p_{\gamma_i}})\cdot \widetilde{ \cI\!\!\cI }
- \sum_{i=1,2} t^3_p (1 - t^3_{p_{\gamma_i}})\cdot \widetilde{ \cI\!\!\cI }\,.
\label{W3B7}
\ee
The contributions related to $\gamma_1$ and  $\gamma_2$, namely the
last two terms in \bref{W3B7}, are identically zero by exactly
the same argument that was given in the previous subsection
(cf.\,eq.\,\bref{int63}). The difference in
eq.\,\bref{sigma dif} is
\be
   \Sigma_2(x-y)- \Sigma_0(x-y)=
   2 i \hbar^3 d^2  \int {\dif^2 p \over (2\pi)^2}\,\,
   e^{ i p \cdot (x-y)} {p \over p\bar p + m^2} \,\,
   \widetilde{ I\!\!I}_R(p,m),
\label{W3B8}
\ee
with the 1PI part given
by
\bea
   \widetilde{ I\!\!I}_R(p,m) &=&
   \int {\dif^2 l \over (2\pi)^2}\,
   \left( t^3_p - t^5_p \right)
   {(p-l)\over \left[(p-l)(\bar p - \bar l)+m^2 \right]}\times
\nonumber\\
   &&\int {\dif^2 k \over (2\pi)^2}\,
   \left( 1 - t^2_l \right)\left[ {k^2 \over k \bar k + m^2}\,
   {(l-k)^2 \over \left[ (l-k) (\bar l-\bar k)+ m^2\right]}\right]\,.
\label{W3B9}
\eea
The same integration variables as
in \bref{int66} were used.
The  consequence of the anisotropic normal
product is that the subtraction of the one loop subdivergences is
the minimal one. For the overall subtraction, there is a difference
and the entire two-loop anomaly is due to this.

We now compute
$\widetilde{ I\!\!I}_R$. Using
\vgl{MasRen1Lint} for the one-loop, minimally subtracted $k$ subintegral,
we obtain
$$
  \widetilde{ I\!\!I}_R(p,m) = - {i \over 2\pi} \int_0^1 dx \, x^2 (1-x)^2
  \int {\dif^2 l \over (2\pi)^2}  \,\,
   \left( t^3_p - t^5_p \right)
   {(p-l) \over \left[(p-l)(\bar p - \bar l)+m^2 \right]}
   {\left[l^4 \over x(1-x) l \bar l + m^2\right]}\, .
$$
Only the terms of the form $(l\bar l)^n$ in
the difference between the Taylor series  survive
the $l$ angular integral, such that \bref{W3B9} becomes
$$
  \widetilde{ I\!\!I}_R(p,m) =
  {i \over 2\pi} p^5 m^2  \int_0^1 dx \, x^2 (1-x)^2
  \int {\dif^2 l \over (2\pi)^2}  \,\,
  {\left[1 \over x(1-x) l \bar l + m^2\right]}
  { (l\bar l)^4 \over ( l\bar l + m^2 )^6 } \, .
$$
The Wick rotated $l$ momentum integral is conveniently performed
using  polar coordinates, and yields after changing
radial integration variable to $z = \vert l \vert^2 / 2 m^2$,
$$
  \widetilde{ I\!\!I}_R(p,m) =  {1 \over 4 \pi^2} {p^5 \over m^2}
 \int_0^1 dx \, x^2 (1-x)^2 \,\, \int_0^{\infty} dz \,
 {z^4 \over (z + 1)^6 [ x(1-x) z + 1]} \, .
$$
A further change of variables $z = t / (1-t)$, followed by the
change to \\
$\{\alpha_1  = tx, \alpha_2 = t(1-x), \alpha_3 = 1-t\}$ brings it in the
more familiar form
\bea
    \widetilde{ I\!\!I}_R(p,m) & = &
   {1 \over 4 \pi^2} {p^5 \over m^2}
   \left( \prod_i \int_0^1 d \alpha_i \right) \delta\left( \sum_{i=1}^3
   \alpha_i - 1  \right) \,\, {\alpha_1^2 \alpha_2^2 \alpha_3
   \over  \alpha_1 \alpha_2 + \alpha_1 \alpha_3 +  \alpha_2 \alpha_3}
\nonumber \\
  & = &  {1 \over 4 \pi^2} {p^5 \over m^2} \frac13
  \left( \prod_i \int_0^1 d \alpha_i \right) \delta\left( \sum_{i=1}^3
  \alpha_i - 1  \right) \,\, \alpha_1 \alpha_2 \alpha_3  \, ,
\nonumber
\eea
where the last line is obtained by symmetrising the previous numerator
in the $\alpha_i$. Further use of the well-known result
$$
  \left( \prod_{i=1}^n \int_0^1 d \alpha_i \right)
  \delta\left( \sum_{i=1}^n \alpha_i - 1  \right) \,\,
  \prod_{i=1}^n  \alpha_i^{a_i - 1} =
 { \prod_{i=1}^n \Gamma (a_i) \over \Gamma(\sum_i a_i)},
$$
gives
$$
    \widetilde{ I\!\!I}_R =
    {1 \over 2 \pi^2} {p^5 \over m^2} {1 \over 6!}\, .
$$
When plugged back in (\ref{sigma dif}, \ref{W3B8}) this determines the
coefficient $\rho$ in the Zimmerman identity \bref{w3zid}:
\be
  \rho=-\frac{d^2\hbar^2}{6! \pi m^2}.
\label{rhow3}
\ee
Insertion of this Zimmerman identity in eq.\,\bref{W3B1} leads then to the
cancellation of its overall $m^{2}$ factor with the $m^{-2}$ factor in
\bref{rhow3} and finally gives a nonvanishing two-loop contribution to
the anomaly \bref{W2B1}
in the limit $m \rightarrow 0$. We find
$$
   \cA_2 = {i \hbar d^2 \over 720 \pi^2} \int \dif ^2x \,\,
    u \,\partial^5 v^* \, ,
$$
which is in complete agreement with the previous result \bref{an2}.

\section{Conclusions and Outlook}

\hspace{\parindent}%
In this paper, we have started a program to put the quantum theory of the
BV quantisation formalism on a sounder basis. The old machinery of
BPHZ renormalisation and normal ordered products turns out to be well
suited for this purpose. We have given a formulation of the quantum
BV master equation using these normal ordered products, which is valid
{\it to all orders in perturbation theory}.
An advantage over the Zinn-Justin formulation in terms of the effective
action is that we work with {\em local} operator equations.
As a concrete testing ground, we have computed the BRST
anomalies for some two dimensional field theories, where it is known that,
in contrast with Yang-Mills theories, new anomalies appear in higher loops.

Apart from the possibility to compute the anomaly --
to all orders in perturbation theory --
directly from the Veltman-Ward identity, a closed
local expression for the anomaly was obtained (\vgl{anomeq2}). Two
ingredients are of crucial importance here. First,
to apply the BPHZ method to massless theories one should temporarily
introduce a mass term: the anomaly is related to the
explicit breaking of the BRST symmetry by this mass term. This is
reminiscent of, but different from, the one-loop Pauli-Villars method.
The second ingredient is the use of anisotropic
oversubtracted normal ordered operators.
They originate in the mass term, and occur
with a coefficient proportional to
the mass (squared). The massless limit does not vanish however: it
is extracted by rewriting the
oversubtracted operator in terms of a set of minimally subtracted ones,
using Zimmerman's normal product identity. The anomaly is what remains in
this limit.

We have successfully applied our method to the computation of the
anomalies in $W_2$ and $W_3$ gravity. In particular, detecting the
presence of the two-loop anomaly of the latter model is as easy as
detecting the one-loop anomaly of either model. Only the computation of
the coefficient, involving the renormalisation of a two-loop diagram
using forest formula techniques, is more involved.

Of course, these methods can also be applied
to compute anomalies in global symmetries. As an example,
we verified that with our choice of
mass term, the correct \gn anomaly is reproduced. Despite the fact
that both the original action and the mass term have \gn zero, the
associated Noether current is {\it not} conserved, owing to a
non-locality in the mass term, and again the occurrence of oversubtracted
operators.

Having demonstrated the possibilities offered by the BPHZ method in
setting up proper renormalised versions of the quantum master equation, we
now give some possible directions for further development.

A straightforward extension would be to include
extra finite ``counterterms'' in the action.
In specific examples, these harbour the possibility
to cancel or shift the anomalies of the quantum theory. These
are related to the possibility that the
interaction terms  and/or the BRST transformation rules
are not minimally subtracted, and
presumably also to the choice of mass term, as in
Pauli-Villars regularisation. Also, whereas our local anomaly equation is
in principle valid irrespective of the closing of the algebra off-shell,
it could be of interest to work out explicitly an example where the
extended action is non-linear in the antifields.

On a more theoretical level the question of consistency conditions for
higher loop anomalies imposes itself.
For the effective action we have that
$ ( \Gamma , \cA\cdot\Gamma ) = 0 \, . $
Formally, it is easy to derive the corresponding local
conditions perturbatively in $\hbar$,
but unfortunately this is again ill-defined due to the use of
the formal operator $\Delta$ (see section~\ref{BV}).
To lowest order, this corresponds to the condition of Wess and Zumino
that the one-loop anomalies are {\em consistent}, i.e.\ BRST closed.
In the BPHZ formulation that we gave, the offending $\Delta$ is absent%
\footnote{Another possibility is to stay with $\Delta$, but de-localise
all the functionals, as in the nonlocal regularisation scheme treated in
\cite{j95}. See \cite{jw95} for further developments in this direction.}.
The question of higher-loop Wess -Zumino condition is therefore open.
Our local anomaly formula, \vgl{anomeq2},
suggests that the normal product equations could be rather simple
transcriptions of the classical equations, provided one uses the
appropriate anisotropic versions.

The definition of the normal products takes as its starting point a set of
well-defined Feynman rules. This means, in BV, that one starts from a
certain gauge fixed basis. It would be interesting to find a formulation
that does not depend on this choice, or at least to show how
the transition from one choice to another one can be made.
This poses of course the problem of
treating canonical transformations (in the antibracket sense)
on the quantum level
with the BPHZ method. We recall that within the BV scheme, a
change of gauge is implemented via a canonical transformation.
Possible counterterms that are generated by this are also related to
the Jacobian of the BRST transformation. Although
for a closed algebra it is formally independent
of the gauge, in a regularised computation%
\footnote{ We refer to the case of Yang-Mills regularised with
Pauli-Villars, see \cite{SixPack}.}
one may obtain a gauge dependent
expression, owing to the gauge dependence of the regulator.

The final task would be the unraveling of the
quantum cohomology. On the non-local level, the quantum BRST
transformation  can be defined as
$(\Gamma, X)= \cS X$.
For non-anomalous theories, this is a nilpotent operator.
It can be translated back formally to the level of local field theory,
where the quantum BRST operator is $\sigma X = (X, W) - i \hbar \Delta
X$, again involving the ill-defined  $\Delta$. One may hope that the
normal products of Zimmerman
will enable one to define a replacement for this
operation, not just for the action (this is what the anomaly gives), but
also for arbitrary functionals of fields and antifields.

\section*{Acknowledgements}

\hspace{\parindent}%
It is a pleasure to thank Fernando Ruiz Ruiz for discussions on
various aspects of regularization and renormalization.
F.\,D.\,J.\, is supported by the Human Capital
and Mobility Programme through the network on
{\it Constrained Dynamical Systems.}
J.\,P.\, acknowledges financial support from the Spanish
ministry of education (MEC).

\appendix

\section{BPHZ renormalized one-loop integrals}
\label{apint}

\hspace{\parindent}%
Our main goal in this appendix is to compute the general form of the BPHZ
renormalized, minimally subtracted, one--loop integrals used throughout
the paper.

At the one--loop level, the specific form of the interactions for chiral
$W_2$ and $W_3$ gravity
leads to connected Green's functions which always
contain as a 1PI part a special case of the general one-loop integral
$I_{ab}(p,m)$ \bref{gen1Lint}
\be
    I_{ab}(p,m) = \int {\dif^2 q \over (2\pi)^2} { q^a (p-q)^b
    \over (q\bar q+m^2)[(p-q)(\bar p- \bar q) + m^2 ]}\equiv
    \int {\dif^2 q \over (2\pi)^2}\, \cI_{ab}(q,p,m),
\label{1lapp}
\ee
with $ a$, $b \geq 1$.
With a superficial degree of divergence given by
$a+b -2$, minimal subtraction prescribes
that these integrals are replaced by
\be
   I^R_{ab}(p,m)=[(1 - t^{a+b -2}_p)\cdot I_{ab}(p,m)] \, ,
\label{2lapp}
\ee
using the shorthand \bref{shn}.

First one
expresses the  propagators in \bref{2lapp} by means of their well-known
integral representation $a^{-1} = \int_0^\infty dt \, e^{-at}$, i.e.
$$
  I^R_{ab}(p,m)= \int {\dif^2 q \over (2\pi)^2} (1 - t^{a+b -2}_p)
  \int _0^\infty dt_1 \int _0^\infty dt_2 \, e^{-t_1(q\bar q+m^2)}
  e^{-t_2 [(p-q)(\bar p- \bar q) + m^2 ]}
  q^a (p-q)^b \, .
$$
After changing integration variables to
$t = t_1 + t_2$ and $x = t_2 / t$, this becomes
\be
  I^R_{ab}(p,m)= \int {\dif^2 q \over (2\pi)^2}
  \int _0^\infty \dif t \, t \int _0^1 \dif x \,
  e^{-t(q\bar q+m^2)} (1 - t^{a+b -2}_p)
  \left[q^a (p-q)^b\, e^{- tx [ p\bar p - q \bar p - \bar q p]}\right].
\label{3lapp}
\ee

For the actual computation of this integral it is
not necessary to know the subtraction terms in detail, the knowledge of
their general form is sufficient. The Taylor series
expansion around vanishing external momenta of the relevant factor of
the integrand in \bref{3lapp} is
$$
   q^a (p-q)^b\, e^{- tx [ p\bar p - q \bar p - \bar q p]}=
$$
$$
   \sum_{k=0}^b \sum_{n=0}^\infty
   \sum_{r,s=0}^{r+s \leq n}   (-1)^{k+r+s+n} {b \choose k}\,
   {1 \over r!\,s!\,(n-r-s)!}\, t^n x^n \,\,
    \bar q^r \, q^{a+k+s}  \, \bar p^{n-r}\, p^{b-k+n-s}.
$$
The terms that must be subtracted in the minimal scheme are
determined by the inequality
$$
   b-k+2n-r-s   \leq a+b-2,
$$
from which, upon use of $r+s \leq n$, the inequality
$a+k+s >  r$ follows.
All the subtraction terms have therefore
the following $q$ dependence
\be
     q^y (q\bar q)^x,
     \quad\quad\mbox{with}\quad\quad y > 0,
\label{subform}
\ee
i.e. they always contain extra factors $q$, not paired with $\bar q$.
The renormalized, minimally subtracted integral \bref{3lapp} becomes then
\be
  I^R_{ab}(p,m)= \int {\dif^2 q \over (2\pi)^2}
  \int _0^\infty \dif t \, t \int _0^1 \dif x \,
  e^{-t(q\bar q+m^2)}
  \left[q^a (p-q)^b\, e^{- tx [ p\bar p - q \bar p - \bar q p]}
    - {\sum}' a_{r,s} q^y (q\bar q)^x  p^r \bar p^s \right],
\label{subtint}
\ee
where ${\sum}'$ stands for the sum restricted to the terms satisfying the
conditions $r+s \leq a+b-2$, $y >0$. Treated separately, the terms in
the integrand of \bref{subtint} may be UV divergent, but
keeping them together the integral is by construction UV
convergent. The mass $m$ ensures IR convergence
(the region $t\rightarrow\infty$) as well.
In conclusion, \bref{subtint} is an
absolutely convergent multiple integral on which manipulations like
changes of variables, interchanges in the
order of the integration, etc, are mathematically well-defined,
{\it provided} the integrand is treated as a whole.

Now we use this freedom, and  {\it first} perform the
integral over $q$. Due to the general form \bref{subform}
of the subtraction terms and the general result
\be
   \int {\dif^2 q \over (2\pi)^2} e^{-q\bar q} q^n=0,
\label{q0}
\ee
none of them survives the $q$ integration, yielding
\be
   I^R_{ab}(p,m)= \int _0^1 dx \,\int _0^\infty dt \, t \,
   \int {\dif^2 q \over (2\pi)^2} q^a (p-q)^b
   e^{-t [(q-x p)(\bar q - x \bar p) +M(x)]} \, ,
\label{4lapp}
\ee
with $M(x)=m^2 + x(1-x) p \bar p$ and
where now the order of integration is {\it fixed}.
The net effect of the subtraction terms has been that they
have imposed a fixed integration order. Changing
the loop momentum integration variable to $k = q - xp$ and
using \bref{q0}, brings \bref{4lapp} in the simpler form
$$
   I^R_{ab}(p,m)= \int _0^1 dx \,\int _0^\infty dt \, t \,
   \int {\dif^2 k \over (2\pi)^2}
   e^{-t k\bar k} e^{- t M(x)} x^a (1-x)^b p^{a+b} \, ,
$$
which after trivial (Wick rotated) $k$ and $t$ integration (in
{\em this} order) results in
\be
   I^R_{ab}(p,m)=  {-i \over 2\pi} p^{a+b} \int _0^1 dx \,
   { x^a (1-x)^b \over m^2 + x(1-x) p \bar p }=
   I^R_{ab}(p,0)\,F_{ab}\left(\frac{m^2}{p\bar p}\right)\, ,
\label{MasRen1Lint}
\ee
with $F_{ab}(0)=1$. The massless limit provides the desired result
\be
  I^R_{ab}(p,0)={-i \over 2\pi}\, {p^{a+b-1} \over \bar p}
  \int_{0}^{1}  \dif x \,\, x^{a-1}(1-x)^{b-1}
  = {-i \over 2\pi}\, {p^{a+b-1} \over \bar p} B(a,b) \, ,
\label{Ren1Lint}
\ee
in terms of the well-known Euler Beta function $B(a,b)$.
Expression \bref{Ren1Lint} is the general expression for the
minimally subtracted one-loop integrals needed in our computations.

\section{Appendix: The complete one--loop anomaly in $W_3$}
\label{w31lcom}

\hspace{\parindent}%
In this appendix, the computation of the $W_3$ one-loop
anomaly is completed, by
analyzing all the possible double contractions between $R^A(x)$ and
$R^B(y)$ in eq.\,\bref{ord1bis}. The analysis is performed
antifield by antifield, where the antifield in question
determines the $R^A(x)$ to be used on the lhs of \bref{ord1bis} and the
``$\Phi^*_A$-anomaly'' or coefficient $F^A$ of the one-loop anomaly that is
determined. The $b^*$-anomaly is computed in sec.\,\ref{1lw3}.

\subsubsection*{$u^*$--anomaly}

\hspace{\parindent}%
The BRST transformation for $u$, $R^u = [2(\partial c) u- c(\partial u)]$,
can not be contracted twice with any other quantity in the $R^A$'s:
no terms with both a $b$ and a $v$ antighost are contained in them.
Consequently, no term proportional to $u^*$ arises in the anomaly.

\subsubsection*{$c^*$--anomaly}

\hspace{\parindent}%
The only term in the BRST transformation $R^c$ of $c$ that can
give double contractions with other terms in  $R^B$
is the monomial $ 2 k T (\partial u)u$. Moreover, this always goes
via the matter loops studied for the $b^*$-anomaly in the main
text. The direct consequence of this simple fact is that all the Ward
identities that hold
for the matter--loop induced part of the $b^*$--anomaly will hold as well
upon replacing  the antifield $b^*$ with the combination
$ - 2 k c^* (\partial u) u$. In fact, this amounts to rewriting the
relevant interaction term $c^* [2 k T (\partial u)u]$ in \bref{W3action}
as the product of an ``effective'' antifield
$\tilde h^*= - 2 k c^* (\partial u) u$, with exactly the same quantum
numbers as $b^*$, times minus the matter energy momentum tensor $T$.
It is then clear that the $c^*$-anomaly is, in
analogy with \bref{an1b}
$$
   \cA^{c^*}_{1, \rm m} =  { i \over 24 \pi} \int \dif^2 x \,\,  c^{ij}
     \partial^3 [ 2 k c^* u (\partial u) \delta_{ij} ]  \, ,
$$
with $c^{ij}$ defined in (\ref{caaijee}).
The same argument allows a simple determination of the
contributions to the $\phi^*$ and $v^*$ anomalies induced by matter
loops as well.

\subsubsection*{$\phi^*$--anomaly}

\hspace{\parindent}%
In the BRST transformation for the matter fields
\be
R^{\phi^i} =u d^i_{\,jk} (\partial\phi^j) (\partial\phi^k) +
  c(\partial\phi^i) -2 k b (\partial u) u (\partial\phi^i)  \, ,
\label{int37}
\ee
the first monomial,
can only produce contributions via matter loops,
 in analogy with the
$c^*$--anomaly.
Substituting  $b^*\delta_{ij}$ in \bref{an1b} by the effective
antifield $- 2 u \phi^*_k d_{ijk}$ leads then to the following matter
induced contribution to the $\phi^*$ anomaly
$$
  \cA^{\phi^*}_{1,\rm m}=  { i \over 24 \pi} \int \dif^2 x \,\,  c^{ij}
     \partial^3 ( - 2 u \phi^*_k d_{ijk} )  \, .
$$

The remaining two terms in \bref{int37} generate contributions to the
$\phi^*$ anomaly through mixed matter--ghost loops or by purely ghost
loops. Indeed, the monomial $c(\partial\phi^i)$ gives rise
to mixed matter--ghost loops through contractions with the terms
$$
\left\{
   \begin{array}{ccc}
 -2 k b (\partial u) u (\partial\phi^i) & \mbox{ in } & R^\phi \\
 2 k T b (\partial u) + 2 k \partial (Tbu) & \mbox{ in } & R^v \, ,
   \end{array}   \right.
$$
while the monomial $-2 k b (\partial u) u (\partial\phi^i)$ produces
also mixed matter--ghost loops through contractions with the term
$c(\partial\phi^i)$ in $R^\phi$ and purely ghost loops through contractions
with the term $3 v (\partial c) + (\partial v) c$ in $R^v$.
A good test product of fields for all these four possibilities
is $\phi^j(y) v(z) v(t)$, which forces the corresponding anomaly
to contain one $\phi$ field, two $u$ fields and four
derivatives~$\partial$. The one-loop anomalous Ward identity
(\ref{ord1bis}) takes the form
$$
   \begin{array}{c}
  \ihbar \expect{ \norm{0}{c(\partial\phi^i)(x)}
  \norm{0}{-2 k b (\partial u) u (\partial\phi^j)(y)}v(z)v(t)}_{c,0} \\
  \\
  - \ihbar \expect{ \norm{0}{c(\partial\phi^i)(x)}
  \norm{3}{2 k T b (\partial u)(z)+2k\partial(Tbu)(z)}v(t)\phi^j(y)}_{c,0}
    - (z,t) \\
  \\
  + \ihbar \expect{ \norm{0}{-2 k b (\partial u) u (\partial\phi^i)(x)}
  \norm{0}{ c(\partial\phi^j)(y)} v(z) v(t)}_{c,0} \\
  \\
  - \ihbar \expect{ \norm{0}{-2 k b (\partial u) u (\partial\phi^i)(x)}
  \norm{3}{3 v (\partial c)(z) + (\partial v) c(z)} v(t) \phi^j(y)}_{c,0}
    - (z,t) \\
   \\
   = - \alpha_5 \expect{\norm{0}{F^i(\phi,u,u,\partial^4;x)}
       \phi^j(y) v(z) v(t)}_{c,0} \, .
   \end{array}
$$
After a lengthy but straightforward calculation, the lhs is found to be
\bea
   &\displaystyle{
   { k \hbar^3 \over \pi} \int {\dif^2 l \over (2\pi)^2}
   \int {\dif^2 q \over (2\pi)^2}  \int {\dif^2 r \over (2\pi)^2}
    e^{ i (q + l + r) \cdot x} e^{- i q \cdot y} e^{- i r \cdot t}
    e^{- i l \cdot z}}&
\nonumber\\
   &\displaystyle{
    \times {1 \over \bar q \bar l \bar r} \left[ - \frac12 (r-l)(l+q+r)^2
    - \frac12 (r-l) q^2 + 2 (r-l) rl + \frac56 (l^3 - r^3) \right]
    \delta^{ij},}&
\label{int41}
\eea
where the factor $1/\bar q$ corresponds to a
propagator between $\partial \phi$ and $\phi$ and the factors $1/\bar
r$ and $1/\bar l$ are generated by $(u,v)$ ghost propagators.
The combinations of momenta in \bref{int41} can be associated
to the following monomials
$$
    \left\{
  \begin{array}{lcl}
     l^3 - r^3 & \longleftrightarrow & u(\partial^3 u)(\partial \phi) \\
     r^2 l - l^2 r & \longleftrightarrow &
     (\partial u)(\partial^2 u) (\partial \phi) \\
    (r-l)q^2 & \longleftrightarrow & u (\partial u)(\partial^3\phi) \\
   (r-l)(l+q+r)^2 & \longleftrightarrow & \partial^2 \left( u
   (\partial u)(\partial \phi) \right) \, .
  \end{array} \right.
$$
After computing the coefficients of the different
terms in the usual way and performing a few integrations by parts, the
second contribution to the $\phi^*$ anomaly is
\be
   \cA^{\phi^*}_{1,\rm gh}
   = {-i k \over 6 \pi} \int \dif ^2 x \,\, \phi^*_i
\left[ 6 \partial ( u (\partial u)( \partial^2 \phi^i) )
    + 9 (\partial^2 u) (\partial u )(\partial \phi^i) +
    8 u (\partial^3u) (\partial \phi^i) \right]  \, .
\label{an15}
\ee

\subsubsection*{$v^*$--anomaly}
\hspace{\parindent}%
Let us finally compute the contribution to the one-loop anomaly induced by
the BRST transformation of the ghost field $v$
\be
 R^v = -W +2 k T b(\partial u) +2 k \partial(T b u)
       + 3 v(\partial c) + (\partial v) c \, .
\label{rv}
\ee
 The
contributions caused by matter loops
are generated by double contractions between
matter fields in $-W$ or in the energy-momentum tensor $T$ in \bref{rv}
with $(\partial\phi)^2$ type terms in the $R^B$'s.
When writing the
interaction terms in the form of effective sources times $-T$
in analogy with the matter induced $b^*$--anomaly,
the following matter
induced contribution the one-loop $v^*$-anomaly is inferred
$$
   \cA^{v^*}_{1,\rm m} = {i \over 24 \pi} \int \dif ^2 x \,\,
   c^{ij} \partial^3 \left[2 v^* d_{ijk} (\partial \phi^k)+
    2k b ( u \partial v^* - v^* \partial u)\delta _{ij}\right] \, .
$$

The computation of the contribution
of the loops with at least one ghost line, requires more effort.
Closed loops of one matter field and one ghost or of two ghosts can be
obtained in the following four ways.
The terms $[2 k T b(\partial u) +2 k \partial(T b u)]$ get
contracted twice with
$$
 \left\{
   \begin{array}{ccc}
    3 v(\partial c) + (\partial v) c  & \mbox{ in } & R^v \\
    c(\partial\phi^i)  & \mbox{ in } & R^\phi \, ,
    \end{array}
 \right.
$$
while the terms $[3 v(\partial c) + (\partial v) c]$ can contract with
$$
 \left\{
   \begin{array}{ccc}
      -2 k b (\partial u) u (\partial\phi^i) & \mbox{ in } & R^\phi \\
      2 k T b(\partial u) +2 k \partial(T b u)& \mbox{ in } & R^v \, .
   \end{array}
\right.
$$
A suitable testproduct for all four contributions is $\phi^j(y) \phi^k(z)
v(t)$, yielding a contribution to the $v^*$-anomaly
constructed out of two $\phi$-fields, one $u$ ghost
and five derivatives $\partial$, and a one-loop anomalous Ward identity
of the form
\be
   \begin{array}{c}
  - \ihbar
  \expect{\norm{3}{2 k T b(\partial u)(x) +2 k \partial(T b u)(x)}
\norm{3}{3 v(\partial c)(t) +(\partial v)c(t)}\phi^j(y)\phi^k(z)}_{c,0} \\
  \\
  + \ihbar \expect{\norm{3}{2 k T b(\partial u)(x) +2 k \partial(T b u)(x)}
    \norm{0}{c(\partial\phi^j)(y)} \phi^k(z) v(t)}_{c,0} + (j,k)(y,z) \\
  \\
  + \ihbar \expect{\norm{3}{3 v(\partial c)(x) + (\partial v) c(x)}
  \norm{0}{-2 k b (\partial u) u (\partial\phi^j)(y)} \phi^k(z) v(t)}_{c,0}
     + (j,k)(y,z) \\
  \\
   - \ihbar \expect{\norm{3}{3 v(\partial c)(x) + (\partial v) c(x)}
   \norm{3}{2 k T b(\partial u)(t) +2 k \partial(T b u)(t)} \phi^j(y)
   \phi^k(z)}_{c,0} \\
   \\
   = - \alpha_{6,7} \expect{\norm{3}{F(u,\phi,\phi,\partial^5;x)}
   \phi^j(y) \phi^k(z) v(t)}_{c,0} \, .
   \end{array}
\label{int46}
\ee

The equality of the first and fourth contribution on the lhs of
\bref{int46} upon exchanging $x$ and $t$, and the fact that the
value of the second and third contributions can be read off from some of
the results obtained in the $\phi^*$- computation above,
reduce the amount of work necessary to obtain the renormalized
expresion for the lhs of (\ref{int46}). It reads
\bea
   &\displaystyle{
   {- k \hbar^3 \over \pi} \int {\dif^2 l \over (2\pi)^2}
   \int {\dif^2 q \over (2\pi)^2}  \int {\dif^2 r \over (2\pi)^2}
    e^{ i (q + l + r) \cdot x} e^{- i q \cdot y} e^{- i r \cdot t}
    e^{- i l \cdot z}}&
\nonumber\\
   &\displaystyle{
   {1\over \bar q \bar l \bar r} \left[ \frac12 \left( r(l^2 + q^2) +
   (l + q + r) (l^2 + q^2) \right) \right. }&
\nonumber\\
   &\displaystyle{
    \left.   + \frac16 \left( 5 r^3 + 12 r^2
   (l + q + r) + 12 r (l + q + r)^2 + 5 (l + q + r)^3 \right) \right]}&
   \, .
\label{int47}
\eea
It is now a piece of cake to determine
the distribution of the five derivatives $\partial$ in
the coefficients $F(u,\phi,\phi,\partial^5;x)$. Indeed, taking into account
that the factor $(\bar q \bar l \bar r)^{-1}$ is
generated by two propagators between $\phi$ and $\partial \phi$, and one
propagator $(u,v)$, the other momentum functions can be associated
with the following terms in $F(u,\phi,\phi,\partial^5;x)$
$$
 \left\{
  \begin{array}{lcl}
     r (l^2 + q^2)  & \longleftrightarrow &
     (\partial \phi) (\partial^3 \phi) \partial u \\
     (r + l + q)(l^2 + q^2) & \longleftrightarrow &
     \partial\left( u (\partial \phi) (\partial^3 \phi)\right) \\
     r^3  & \longleftrightarrow & (\partial^3 u )( \partial \phi)
     (\partial \phi) \\
     r^2 (r + l + q) & \longleftrightarrow &  \partial\left(
     (\partial^2 u) (\partial \phi)(\partial \phi) \right) \\
     r(r + l + q)^2 & \longleftrightarrow &  \partial^2\left(
     (\partial u) (\partial \phi)(\partial \phi) \right) \\
     (r + l + q)^3 & \longleftrightarrow &  \partial^3 \left(
     u  (\partial \phi)(\partial \phi) \right) \, .
  \end{array}
\right.
$$
By simple comparison with (\ref{int47}), the coefficients of
each of these terms are determined and the mixed and ghost
induced contributions to the one-loop $v^*$-anomaly is
\bea
  \cA^{v^*}_{1, \rm gh}& =&
  {i k \over 2 \pi} \int \dif^2 x\,\, \left( v^* (\partial u)
        - u (\partial v^*) \right) ( \partial \phi^i)
        (\partial^3 \phi^i)
\nonumber\\
   &&+ {i k \over 6 \pi} \int \dif^2 x\,\, T \left[ 5 (\partial^3 u) v^*
   - 12 (\partial^2 u) (\partial v^*)  + 12 (\partial u )
   ( \partial^2 v^*) - 5 u (\partial^3 v^* )\right] \, .
\label{an19}
\eea

In summary, collecting all the previous contributions,
we can write the complete one-loop anomaly for $W_3$ in the compact way
\bref{c1lw3}. This finishes our analysis of the one-loop contribution to
the BRST anomaly in the $W_3$ model.

\section{The ghost number anomaly}
\label{ghanom}

\hspace{\parindent}%
In this appendix, we show how the well-known \gn anomaly \cite{Kazuo}
is obtained in the context of the BPHZ scheme. Although in
principle this is another application of the method sketched in
sect.\,\ref{Aandmass} for computing the global axial anomaly, we
find it instructive to present it as an example of how
nonlocal mass terms, though invariant, may give rise to
anomalous conservation laws. Also, the fact that the correct \gn anomaly
is obtained should dispell any misgivings about the use of
this non-local mass-term.

 In order to make contact with standard
formulations of chiral gravities, we consider a
simplified version of the models used before,
with all antifields set to zero except
the antifield $b^*$, which takes over the r\^ole of (minus)
the gravitational gauge field, i.e.
\be
   \tilde S=\int\dif^2 x
   %\hspace{-5mm}&&
   \left[ -\frac12(\partial\phi)(\bar\partial\phi)+
    b_j(\bar\partial c_j) -{m^2\over 2} \left( \phi^2 +
   2 b_j {1 \over \partial} c_j \right)
   -b^*\left[T +T_2+T_3\right]\right] \, .
\label{massive restricted}
\ee
$\{c_j, b_j\}$, $j=2,3$ are the spin 2 and 3 ghost-antighosts
systems and $T_2$, $T_3$ their corresponding energy momentum tensors
\bref{Teejay}. As it stands, this  massive action \bref{massive restricted}
is {\em invariant} under the following ghost number transformations
\be
   \delta_\sharp \Phi^A= \sharp_A \Phi^A,\quad\quad
   \delta_\sharp b^*= 0,
\label{gh trans}
\ee
where the ghost numbers $\sharp_A$ for the fields
are listed in tables \ref{tbl:charges} and \ref{tbl:chargesII}.
One would naively conclude that there is no room for a ghost
number anomaly in our formulation, contrary to the well-known
results. However, this is not the
case and, once again, the oversubtractions performed on the IR
regulating mass term, together with its nonlocal character, are the
origin of the \gn anomaly.

As usual for global symmetries, the analysis starts by computing
the divergence of the ghost number current
$\left( S = \tilde S(m=0) \right)$
\be
   \partial_{\mu} j^{\mu}(x) =
   -\left[ \frac{\dr S}{\delta \Phi^A(x)} \delta_\sharp \Phi^A(x)\right]
   +m^2 \sum_{j=2,3}
   \left[\left({1 \over \partial} b_j\right) c_j +
         b_j  \left({1 \over \partial} c_j\right) \right] \, ,
\label{g4}
\ee
which classically vanishes on--shell in the limit $m \rightarrow 0$.
The anomalous Ward identity is
\bea
   &\displaystyle{
  \expect{ \norm{2}{\partial_\mu j^{\mu}(x)}
  \prod_i \Phi^{A_i}(x_i) }_c =} &
\nonumber\\
   &\displaystyle{
- i \hbar \sum_i \sharp_{A_i}
   \expect{ \prod_i \Phi^{A_i}(x_i) }_c \delta ( x - x_i ) +
   \expect{ \norm{2}{\cG(x)} \prod_i \Phi^{A_i}(x_i) }_c, }&
\label{g5}
\eea
where the operator insertion $\cG(x)$ parametrises the \gn anomaly.
The linearity of the \gh transformations \bref{gh trans}
allows, upon application of the Schwinger-Dyson equations
\bref{ani SD-BPHZ}, to rewrite the contribution from the massless
piece in \bref{g4} as
$$
  \expect{\norm{2}
  {-\left( \frac{\dr S}{\delta \Phi^A(x)} \delta_\sharp \Phi^A(x)\right)}
  \prod_i \Phi^{A_i}(x_i) }_c=
 - i \hbar \expect{
 \left(\delta_\sharp \Phi^A(x) \frac{\dl }{\delta \Phi^A(x)} \right)
 \prod_i \Phi^{A_i}(x_i) }_c ,
$$
which reproduces the first, non-anomalous term on the rhs of the
Ward identity \bref{g5}. The \gn anomaly is therefore
determined by the {\it oversubtracted} operator
$$
  m^2\sum_{j=2,3}\norm{2}{ \left({1 \over \partial} b_j\right) c_j(x) +
  b_j\left({1 \over \partial} c_j(x)\right)}.
$$
No anisotropic normal orderings appear because the transformation rules of the
\gn symmetry are linear.

Once again, prior to taking the limit $m\rightarrow 0$,
the Zimmerman identity \bref{ZimId} should be used
\be
   m^2\sum_{j=2,3}\left\{
   \norm{2}{ \left({1 \over \partial} b_j\right) c_j +
    b_j\left({1 \over \partial} c_j\right)}-
   \norm{0}{ \left({1 \over \partial} b_j\right) c_j +
    b_j\left({1 \over \partial} c_j\right)}\right\}
    = \sum_k r_k \norm{2}{\cG_k} \, .
\label{g8}
\ee
Its rhs is determined in the usual way by inserting
both sides of this relation in suitable correlation functions.
In doing this computation, however, some work can be saved by
using the dimensional arguments presented in
sect.\,\ref{w2awi}. They show that the insertions
$\cG_k(x)$ in \bref{g8} must be of the form
$\cG_k\sim G(\partial) b^*$. In this way, it is
concluded that the one-loop contributions to the correlation functions
\bea
  &\displaystyle{ M^j_a (x)
  = \ihbar \int \dif^2 y \, b^*(y) I^j_a(x-y)= }&
\nonumber\\
  &\displaystyle{ \ihbar \int \dif^2 y \, b^*(y)
  \expect{ \norm{2}{ j b_j(\partial c_j)+ (j-1)(\partial b_j) c_j}(y)
 \norm{a}{\left({1 \over \partial} b_j\right) c_j +
            b_j\left({1 \over \partial} c_j\right)} }_{c,0}}&,
\nonumber
\eea
given in terms of the BPHZ renormalized one-loop integrals
\be
  I^j_a(x-y) = -\hbar^2 \int {\dif^2 p  \over (2\pi)^2} \,\,
  e^{ i p \cdot (x-y)} \,\, \int {\dif^2 q \over (2\pi)^2}
  (1-t_p^a)\left[
  {pq+(j-1)p^2 \over(q\bar q+m^2)[(p-q)(\bar p- \bar q) + m^2]}\right]\,,
\label{g12}
\ee
provide the relevant information on what the rhs of \bref{g8} exactly is
in terms of the difference $\sum_{j=2,3}( M^j_2(x) - M^j_0(x))$.

Computing the
difference $I^j_2(x-y)-I^j_0(x-y)$ requires the knowledge of the $p$
expansion of the integrand in \bref{g12} up to second order. Keeping
only terms of the form $(q\bar q)^n$ in the Taylor
series  (see appendix~\ref{apint}), we obtain
\bea
  &\displaystyle{ I^j_2(x-y)-I^j_0(x-y)=}&
\nonumber\\
  &\displaystyle{
  \hbar^2 \int {\dif^2 p  \over (2\pi)^2} \,\,
  e^{ i p \cdot (x-y)} \,\, \int {\dif^2 q \over (2\pi)^2}\,
  t_p^2\,\left[
  {pq+(j-1)p^2 \over(q\bar q+m^2)[(p-q)(\bar p- \bar q) + m^2]}\right]=}&
\nonumber\\
  &\displaystyle{
-\hbar^2 \int {\dif^2 p  \over (2\pi)^2} \,\,
  e^{ i p \cdot (x-y)}\,\, p^2\,\, \int {\dif^2 q \over (2\pi)^2}
  \left[{(j-1)\over(q\bar q+m^2)^2}
  +{q\bar q\over(q\bar q+m^2)^3}\right]=}&
\nonumber\\
  &\displaystyle{
\frac{i(2j-1)\hbar^2}{4\pi m^2}
\int {\dif^2 p  \over (2\pi)^2} \,\, e^{ i p \cdot (x-y)}\,\, p^2=
   -\frac{i(2j-1)\hbar^2}{4\pi m^2} \partial^2 \delta(x-y) \, .}&
\nonumber
\eea
Plugging this result back in the  difference
 $\sum_{j=2,3}( M^j_2(x) - M^j_0(x))$,
we easily obtain
$$
 M^j_2(x) - M^j_0(x) =  \frac{(2j-1)\hbar}{4\pi m^2}
 \partial^2 b^*(x) \, .
$$

The Zimmerman identity \bref{g8} then takes the form
$$
   m^2\sum_{j=2,3}\left\{
   \norm{2}{ \left({1 \over \partial} b_j\right) c_j +
    b_j\left({1 \over \partial} c_j\right)}-
   \norm{0}{ \left({1 \over \partial} b_j\right) c_j +
    b_j\left({1 \over \partial} c_j\right)}\right\}=
   \frac{(3+5)\hbar}{4\pi}
    \partial^2 b^*,
$$
from which, upon taking the limit $m \rightarrow 0$, the expression of the
anomalous insertion in \bref{g5} is finally determined
$$
  \cG(x)= \frac{2\hbar}{\pi} \partial^2 b^*(x).
$$
This is in agreement with previous computations in the literature
\ct{Kazuo}.

We conclude that, despite the invariance of the mass term, the correct
anomalous Ward identity is obtained. The anomaly arises from
oversubtracted operators which occur with a coefficient proportional to
the mass (squared). The massless limit of these combinations is extracted
using Zimmerman's normal product identity.

\endsecteqno

\end{document}